\documentclass[twocolumn,english,prx,showpacs]{revtex4-1}
\usepackage[latin9]{inputenc}
\usepackage{amssymb}
\usepackage{graphicx}
\usepackage{amsmath}
\usepackage{color}
\usepackage{float}
\usepackage{mathrsfs}
\usepackage{indentfirst}
\usepackage{textcomp}
\usepackage{comment}
\usepackage{mathtools}
\usepackage{subfigure}
\usepackage{placeins}    
\usepackage[hidelinks]{hyperref}

\newcommand{\ket}[1]{\left| #1 \right>} 
\newcommand{\bra}[1]{\left< #1 \right|} 
\newcommand{\braket}[2]{\left< #1\vphantom{#2}|#2\vphantom{#1} \right>} 

\newcommand{\COMMENTED}[1]{}

\begin{document}

\title{A numerically exact study of Weyl superconductivity}

\author{Peter Rosenberg}
\affiliation{National High Magnetic Field Laboratory and Department of Physics, Florida State University, Tallahassee, Florida 32306, USA}
\author{Niraj Aryal}
\affiliation{National High Magnetic Field Laboratory and Department of Physics, Florida State University, Tallahassee, Florida 32306, USA}
\author{Efstratios Manousakis}
\affiliation{National High Magnetic Field Laboratory and Department of Physics, Florida State University, Tallahassee, Florida 32306, USA}

\begin{abstract}
We study the interplay of interactions and topology in a pseudo-spin Weyl system, obtained from a minimally modified Hubbard model, using the numerically exact auxiliary-field quantum Monte Carlo method complemented by mean-field theory. 
We find that the pseudo-spin plays a key role in the pairing mechanism, and its effect is reflected in the structure of the pairing amplitude. An attractive on-site interaction leads to pairing between quasiparticles carrying opposite spin and opposite
topological charge, resulting in the formation of real-spin singlet pairs that are a mixture of pseudo-spin singlet and pseudo-spin triplet. 
Our results provide a detailed characterization of the exotic pairing behavior in this system, and represent an important step towards a more complete understanding of superconductivity in the context of topological band structures, which will help guide searches for topological superconductivity in real materials and ultracold atoms.
\end{abstract}

\maketitle

\section{Introduction}
Pairing and superconductivity have been focal points of condensed matter physics for several decades, whereas the comparatively modern
discovery of topological materials has generated intense recent interest in the role of topology in condensed matter systems.
The microscopic, many-body origins of superconductivity are the subject of an expansive body of theoretical and computational work, while exhaustive schemes have been developed to classify topological systems based on symmetries of the Hamiltonian and band structure. There is a fundamental conceptual division between these efforts; pairing is a many-body behavior that only emerges in the presence of electron interactions, while topological behavior is typically characterized by the non-interacting one-body picture. A successful unification of these concepts is a crucial step towards a more complete treatment of many intriguing problems, including topological superconductivity. Given the enticing potential applications of these ideas in quantum computing and information, a detailed, quantitative description is a compelling open challenge.

Since the inception of the Hubbard model, quantum lattice models have served as a testbed for theories of pairing and superconductivity.
One prominent example being the repulsive Hubbard model, which has been at the center of a persistent and ongoing effort to understand the origin and mechanisms of superconductivity in the cuprates. These models are broadly relevant across condensed matter, nuclear, and atomic physics, and most recently in the context of ultracold atoms, which offer the possibility of high-accuracy and finely tunable experimental realizations of a variety of lattice models \cite{Cheuk1260,PhysRevLett.111.185302}.

While pairing has been a central theme of condensed matter research for over sixty years, the last decade has seen the emergence of new classes of materials with topological character, in particular, the class of Weyl semimetals, that have revolutionized condensed matter physics. 
These materials are especially interesting because they become superconducting at sufficiently low temperature~\cite{Guguchia2017,MoTe2Rhodes_PRB2017,EnhancedTcMoTe2Rhodes_2019}. The presence of superconductivity in these Weyl systems prompts an intriguing question: to what extent is the superconductivity due to Weyl fermions that seem to be present near the Fermi energy?

Here we introduce a simple lattice fermion model, which is a straightforward extension
of the two-dimensional (2D) Hubbard model, with Weyl quasipartlcles near the Fermi surface that interact via on-site attraction. Other models of Weyl systems have been studied either without interactions \cite{Zhang2016} or at the mean-field level 
\cite{WeylSuperconductivityBalents, WeylSuperconductivityBurkov},   
but these mean-field approaches are approximative and their results can be unreliable. Exact diagonalization, while free of 
approximations, is limited to very small sized systems, and is therefore unsuitable for the detection of 
long range correlations and pairing. Quantum Monte Carlo methods, based on the truly underlying 
degrees of freedom, the electrons, have demonstrated unique capability in the treatment of strongly-correlated
many-body systems, including systems with fermionic pairing \cite{2DFG_PRA,2DFG_SOC_AFQMC,AFQMC_Rashba_OPLATT,AFQMC_RASHBA_INVITE,ettoreGAP}. However,
many lattice models with topological character suffer from the fermion sign problem, so very few  
unbiased many-body studies of strongly-correlated topological systems have been done 
\cite{Sorella-2018,Yao-2017,Fiete-2014,Assaad-2011}, and a quantitative description of pairing in these systems 
remains an important goal. 

With a minimal modification to the well-known attractive Hubbard model,
we obtain a Hamiltonian whose one-body term describes a Weyl system, even in the absence of 
spin-orbit coupling. The one-body part is a set of stacked 1D SSH chains \cite{SSH}
coupled via an inter-chain hopping term to form a 2D lattice.   
The addition of interactions, which render the model analytically intractable and
exceptionally computationally challenging, leads to the emergence of several exotic
behaviors including pairing between Weyl quasiparticles.
We find that with suitable hopping parameters the system supports a phase
with pairs composed of quasiparticles from the same pseudo-helicity branch, but of opposite topological charge (chirality). 
We study the model using the cutting-edge auxiliary-field quantum Monte Carlo (AFQMC) 
method, complemented by mean-field theory calculations, for a variety of system sizes to obtain
good control over the finite-size effects. The AFQMC and mean-field approaches are in reasonable agreement.
The Hamiltonian preserves time reversal symmetry, which guarantees that our AFQMC calculations are free of 
the sign problem \cite{nosignproblem}, meaning that these results are numerically exact and offer a uniquely
accurate, detailed quantitative description of pairing behavior in a strongly-correlated topological system.
In addition to demonstrating the capability of the AFQMC technique to treat interacting
topological systems, this work can serve as a guide to ongoing experimental efforts in topological superconducting materials. 

The quantitative picture we develop yields several observations about the nature
of pairing in this topological model. We find that the attractive interaction leads to spin
singlet pairing of Weyl quasiparticles of the same helicity but opposite topological charge into a BCS-like 
superconductivity, which emerges at a small but finite interaction strength.
We also examine in detail the pairing mechanism and its effects on the
spatial variation of the pairing amplitude. While the constituents of 
each pair have non-trivial topological character, the attractive on-site interaction leads to
a topological-charge-neutral form of superconductivity. We also discuss pairing mechanisms
capable of producing topological superconductivity in this Weyl system \cite{Haldane-2018}, which is a direction of future research.

The remainder of the paper is organized as follows,
In section \ref{sec:Model} we introduce our lattice model.
In section \ref{sec:Methods} we describe our approach, which
combines AFQMC and mean-field theory calculations.
We present our results in section \ref{sec:Results}, and conclude with
a discussion of experimental implications, routes towards topological superconductivity, and an outlook in section \ref{sec:Discussion_Conclusion}.

\section{The Model}
\label{sec:Model}

\begin{figure*}[ht]
  \begin{center}
    \includegraphics[width=0.95\textwidth]{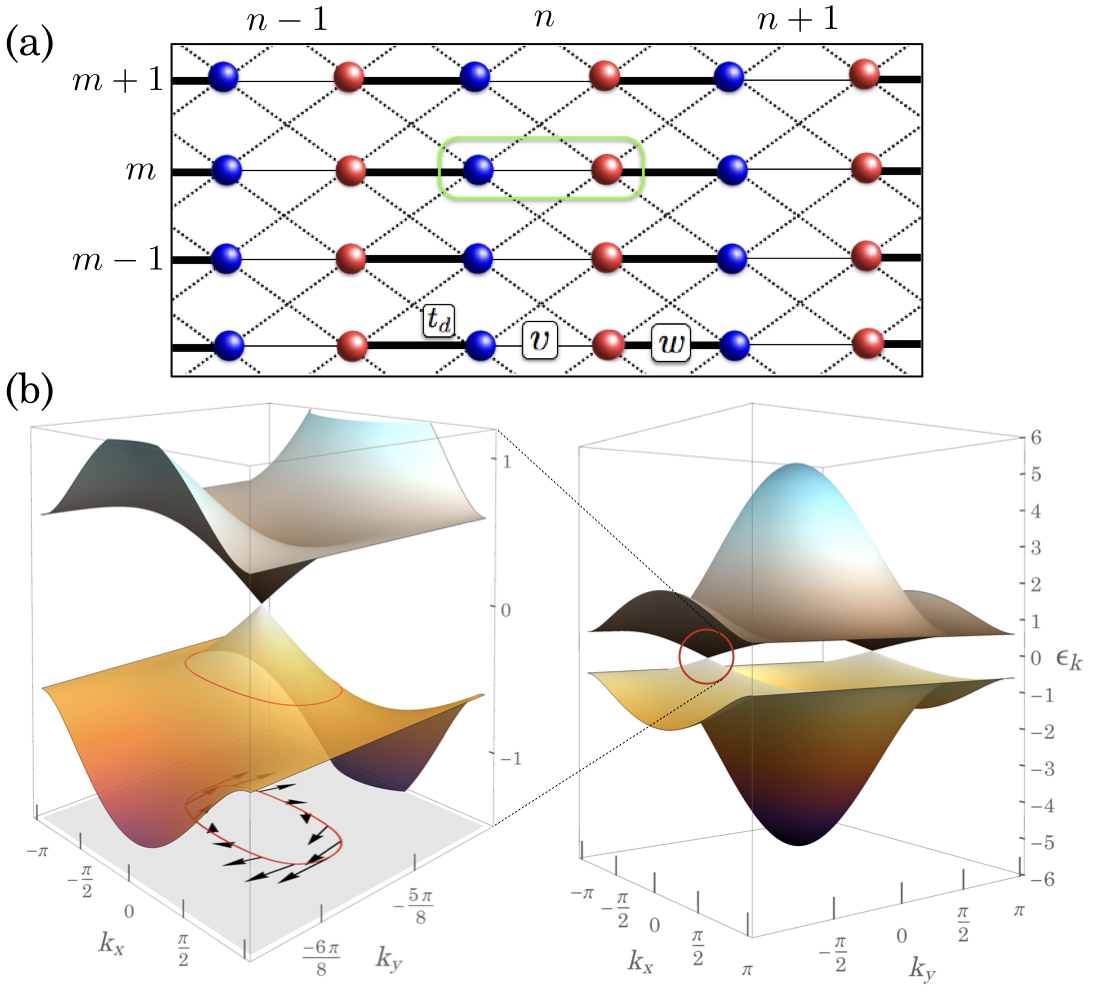}
  \end{center}
        \caption{Lattice  and  band  structure.  In (a)  we  show  the
          geometry of  the lattice  with a  two-site unit  cell (green
          box) labeled by $n$ and $m$. Hopping only occurs between the
          two  different sublattices  (indicated by  the color  of the
          site). The band-structure of the non-interacting Hamiltonian
          is  plotted in (b). We highlight  one of
          the nodal points, characterized by a linear band
          crossing. The red curve along the lower band represents the
          Fermi surface  at a filling  of $n\sim 0.95$, and  the arrows
          show the Berry potential at the Fermi surface. The Berry phase
          around either node is $\pm \pi$, indicating that the nodes carry
          equal but opposite topological charges. This is reflected by the
          vortex behavior of the Berry potential, which has equal magnitude
          but opposite chirality around either node.}
        \label{fig:band_structure}
\end{figure*}
Our model has the following lattice Hamiltonian,

\begin{eqnarray}
  \hat{H}_0&=&-v\sum_{n,m,\sigma}(c^{(\textmd{A})\dagger}_{{n,m}\,\sigma}
  c^{(\textmd{B})}_{{n,m}\,\sigma}+\textrm{h.c.})\nonumber\\
  &-&w\sum_{n,m,\sigma}(c^{(\textmd{A})\dagger}_{{n,m}\,\sigma}c^{(\textmd{B})}_{{n-1,m}\,\sigma}+\textrm{h.c.})   \nonumber\\
    &-&t_d\sum_{n,m,\sigma}(c^{(\textmd{B})\dagger}_{{n,m}\,\sigma}
    c^{(\textmd{A})}_{{n,m\pm1},\sigma}+c^{(\textmd{B})\dagger}_{{n,m}\,\sigma}
    c^{(\textmd{A})}_{{n+1,m\pm1}\,\sigma}\nonumber\\
    &+&\textrm{h.c.})\,
\end{eqnarray}
where the operator $c^{(\textmd{A})\dagger}_{{n,m}}$ creates an electron on the A site of the unit cell at position $\mathbf{R}=na\,\hat{x}+mb\,\hat{y}$. Each unit cell is composed of an A site and a B site. The intra- and inter-unit-cell hopping strengths in the $\hat{x}$-direction are controlled by the parameters $v$ and $w$, respectively, and the diagonal hopping strength is given by $t_d$.  

In momentum space the Hamiltonian can be written, 
\begin{equation}
    \hat{H}_0=\sum_{\mathbf{k}\sigma}\mathbf{c}^\dagger_{\mathbf{k}\sigma}\mathcal{H}(\mathbf{k})\mathbf{c}_{\mathbf{k}\sigma}\,,
    \label{eqn:H0}
\end{equation}
with the vector $\mathbf{c}^\dagger_{\mathbf{k}\sigma}=(c^{(\textmd{A})\dagger}_{\mathbf{k}\sigma}, c^{(\textmd{B})\dagger}_{\mathbf{k}\sigma})$, and the matrix $\mathcal{H}(\mathbf{k})=-\mathbf{h}(\mathbf{k})\cdot\boldsymbol{\sigma}$. This has
the form of a Weyl Hamiltonian, where $\boldsymbol{\sigma}=(\sigma_x,\sigma_y,\sigma_z)$ is a vector of the Pauli matrices in the sublattice (or pseudo-spin) basis, and
\begin{equation}\mathbf{h}(\mathbf{k})=
    \begin{pmatrix}
        v_1+w_1\cos(k_xa) \\
        w_1\sin(k_xa) \\
        0
    \end{pmatrix},
    \label{hk}
\end{equation}
with $v_1=v+2t_d\cos(k_yb)$, and $w_1=w+2t_d\cos(k_yb)$.
We plot the lattice geometry and band structure in Fig.~\ref{fig:band_structure}. For $v+w < 4t_d$ there are two Weyl 
nodes at $\mathbf{k}=(0,\pm k_N)$, with $k_N=\cos^{-1}(-(v+w)/4t_d)$.
These nodes are protected by the combined inversion (A~$\rightarrow$ B, $\mathbf{k}\rightarrow -\mathbf{k}$) and time-reversal symmetry of the lattice Hamiltonian. For a more detailed characterization of the topological features of the model see Appendix \ref{app-A}.

The connection between the pseudo-spin basis and the diagonal pseudo-helicity basis is defined by the unitary transformation,
\begin{equation}
\left(\chi^{(-)\dagger}_\mathbf{k\sigma}, \chi^{(+)\dagger}_\mathbf{k\sigma}\right) = 
\left(c^{(\textmd{A})\dagger}_{\mathbf{k}\sigma}, c^{(\textmd{B})\dagger}_{\mathbf{k}\sigma}\right)
\frac{1}{\sqrt{2}}\begin{pmatrix}
-e^{-i\theta_\mathbf{h_\mathbf{k}}/2} && e^{-i\theta_\mathbf{h_\mathbf{k}}/2} \\
e^{i\theta_\mathbf{h_\mathbf{k}}/2} && e^{i\theta_\mathbf{h_\mathbf{k}}/2}
\end{pmatrix},
\label{eq:AB_to_PM}
\end{equation}
which introduces a new set of creation operators, $\chi^{(\pm)\dagger}_{\mathbf{k}\sigma}$, with the angle, $\theta_{\mathbf{h}_\mathbf{k}}\equiv\tan^{-1}(h^y_\mathbf{k}/h^x_\mathbf{k})$. In the pseudo-helicity basis the tight-binding Hamiltonian takes
the form:

\begin{eqnarray}
  \hat H_0 &=& \sum_{{\bf k},\sigma,\alpha=\pm} \epsilon^{(\alpha)}_{\bf k} \chi^{(\alpha)\dagger}_{{\bf k}\sigma} \chi^{(\alpha)}_{{\bf k}\sigma},\\
  \epsilon^{(\pm)}_{\bf k} &=& \pm \sqrt{(h^{x}_{\bf k})^2 + (h^{y}_{\bf k})^2}\;.
  \label{eq:ek_pm}
\end{eqnarray}

This simple lattice model with a topological band structure provides an ideal setting in which to study the interplay of topology and superconductivity at the many-body level. With this motivation in mind, we proceed by introducing interactions into the system in the form of an on-site attractive Hubbard term, 
\begin{equation}
\hat{H}_\textrm{I}=\sum_{i,\alpha}Un^\alpha_{i\uparrow}n^\alpha_{i\downarrow},
\label{eqn:HI}
\end{equation}
where $i$ labels the unit cell, $\alpha=\textrm{A},\textrm{B}$, and the interaction parameter $U<0$.
Our full Hamiltonian is now,
\begin{equation}
    \hat{H}=\hat{H}_0+\hat{H}_\textrm{I},
    \label{eq:Htot}
\end{equation}
with $\hat{H}_0$ defined by Eq.~(\ref{eqn:H0}), and $\hat{H}_I$ defined by Eq.~(\ref{eqn:HI}).

\section{Methods}
\label{sec:Methods}

\begin{figure*}[!ht]
	\centering
	\includegraphics[width=0.95\textwidth]{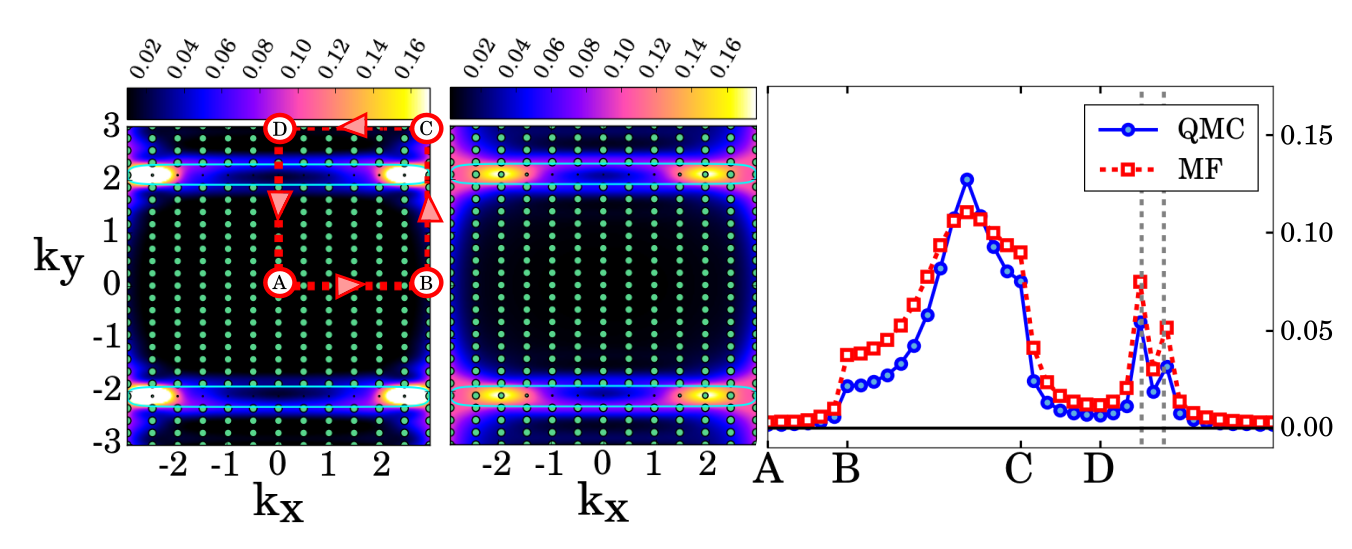}
	\caption{Momentum-space pairing amplitude in the pseudo-helicity basis. The left panel plots $\vert\psi^\textmd{-{}-}_\mathbf{k}\vert$ from AFQMC, and the middle panel plots the same quantity from MFT.  The solid light-blue lines represent the non-interacting Fermi surface, and the size of each green dot is proportional to the total occupation. The right panel shows $\vert\psi^\textmd{-{}-}_\mathbf{k}\vert$ along the path indicated by the red dashed line and arrows in the left panel. In this panel, the non-interacting Fermi surface is indicated by the vertical dashed lines. The system is a periodic $13\times 27$ unit-cell lattice, with $v=0.6$, $w=1.2$, $t_d = 0.9$ and $U=-0.8$.}
	\label{fig:psi_k_mm_QMC_vs_MF}
\end{figure*} 

\begin{figure}[!ht]
	\centering
	\includegraphics[width=\columnwidth]{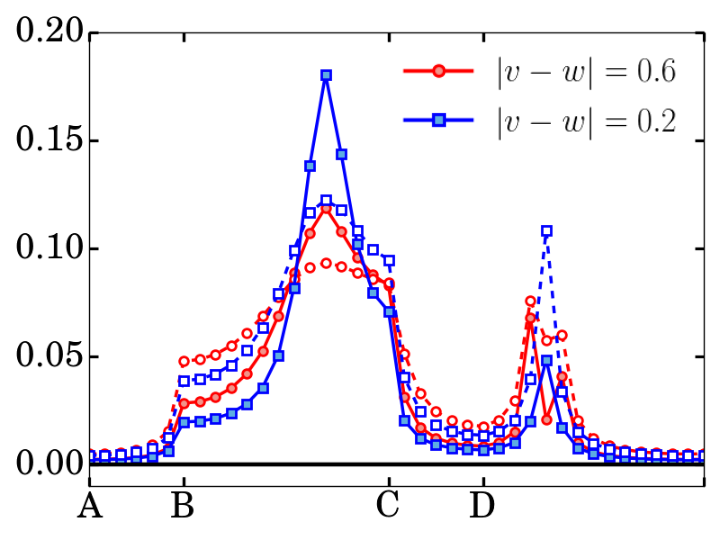}
	\caption{Comparison of momentum-space pairing amplitude from AFQMC and MFT. We plot $\vert\psi^{-{}-}_\mathbf{k}\vert$ for two different values of hopping asymmetry, along the path defined in the left panel of Fig.~\ref{fig:psi_k_mm_QMC_vs_MF}. Results from AFQMC are represented by filled symbols and results from MFT are represented by open symbols. Both systems are periodic $13\times 27$ unit-cell lattices, with $t_d=0.9$, and $U=-1.2$.}
	\label{fig:psi_k_mm_QMC_MF_path2_vs_v-w}
\end{figure}
 
\subsection{Auxiliary-field quantum Monte Carlo}

Strongly correlated many-body systems are a well-known theoretical and computational challenge. One method that has demonstrated considerable accuracy in the treatment of these types of systems is auxiliary-field quantum Monte Carlo (AFQMC) \cite{Koonin,Lecture-notes,AFQMC_Assaad}. The method has been widely applied to both quantum chemistry \cite{Mario_AFQMC_QC,Morales_QMC_Multi} and model Hamiltonians \cite{PhysRevB.94.085103,Ettore_3Band,ettoreGAP,2DFG_PRA,2DFG_SOC_AFQMC,2DFG_RASHBASOC_OPLATT}. In sign-problem-free cases, such as the attractive spin-unpolarized Hubbard model, AFQMC calculations are numerically exact. The technique can be used to calculate an array of ground state properties and provide a reliable quantitative many-body description of the pairing, spin, and charge behaviors of strongly correlated systems.

The AFQMC algorithm is built on the idea of imaginary-time projection, by which the many-body ground-state, $\vert \Psi_0 \rangle$, of a Hamiltonian, $\hat{H}$, can be computed via application of a many-body projection operator to an initial trial wave function, $\ket{\Psi_T}$, 
\begin{equation}
\ket{\Psi_0} \propto \underset{\beta\rightarrow\infty}{\lim} e^{-\beta \hat{H}}\ket{\Psi_T},
\label{eq:proj_lim}
\end{equation}
where $\beta$ is imaginary time, and we require $\braket{\Psi_T}{\Psi_0}\neq 0$ (the trial wave function must have non-vanishing overlap with the many-body ground-state). In the AFQMC framework, the many-body projection operator is decomposed into a set of one-body operators coupled to auxiliary fields. The projection process in Eq.~(\ref{eq:proj_lim}) is then recast as a path integral in auxiliary-field space that can be evaluated using Monte Carlo techniques. See Appendix \ref{app-B} for details on the AFQMC method. Our calculations treat periodic lattices with over 700 sites and 650 electrons to provide a systematic and high-accuracy characterization of the ground-state properties of the system.  

\subsection{Bogoliubov-Valatin-de Gennes-BCS theory}

To complement the AFQMC approach outlined above, we have performed a set of calculations within the BCS mean-field theory (MFT) 
framework. The many-body Hamiltonian in Eq.~(\ref{eq:Htot}) can be written in quadratic form using an appropriate mean-field decomposition,
\begin{eqnarray}
  \hat{H}_{MF} &=& \hat H_0  - \mu \hat{N} + \sum_{{\bf k}\alpha} ( \Delta^{(\alpha\alpha)}  c^{(\alpha)\dagger}_{{\bf k}\uparrow}c^{(\alpha)\dagger}_{-{\bf k}\downarrow}  + \textrm{h.c.}) \,,
\end{eqnarray}
with,
\begin{eqnarray}
  \Delta^{(\alpha\beta)} &=& \frac{U}{N_{c}}\sum_{\bf k^\prime}\langle c^{(\alpha)}_{-{\bf k^\prime}\downarrow}c^{(\beta)}_{{\bf k^\prime}\uparrow}\rangle\,
\label{eq:gap_function}
\end{eqnarray}
where $\alpha,\beta$ take the values A and B, $\mu$ is the chemical potential, $\hat{N}$ counts the total number of electrons, and $N_c$ is the number of unit cells.
The Hamiltonian preserves inversion symmetry (${\bf k} \to -{\bf k}$ and A $\to$~B), which, taking into account the summation over $\mathbf{k}$ in Eq.~(\ref{eq:gap_function}), imposes the following identities:
\begin{eqnarray}
  \Delta^{\textmd{AA}} = \Delta^{\textmd{BB}}, \hskip 0.5 in  \Delta^{\textmd{AB}} = \Delta^{\textmd{BA}}\,.
  \label{eq:deltas}
\end{eqnarray}
In the basis of the helicity eigenstates:
\begin{eqnarray}
  \hat{H}_{MF} &=& \sum_{{\bf k},\sigma,\alpha=\pm} (\epsilon^{(\alpha)}_{\bf k}-\mu)
  \chi^{(\alpha)\dagger}_{{\bf k}\sigma}
  \ \chi^{(\alpha)}_{{\bf k}\sigma} \nonumber \\
  &+& \sum_{{\bf k}\alpha} ( \Delta\chi^{(\alpha)\dagger}_{{\bf k}\uparrow} \chi^{(\alpha)\dagger}_{-{\bf k}\downarrow}   + \textrm{h.c.})\;,
\end{eqnarray}
with $\epsilon^{(\pm)}_{\bf k}$ defined in Eq.~(\ref{eq:ek_pm}) and,
\begin{eqnarray}
 \Delta &=& \Delta^{\textmd{AA}} = \Delta^{\textmd{BB}}.
\end{eqnarray}
The identities in Eq.~(\ref{eq:deltas}) guarantee that pairing only occurs between quasiparticles from the same pseudo-helicity branch. Therefore, in this basis, the mean-field Hamiltonian is block-diagonal, and the gap equation takes the form of the standard BCS gap equation: 
\begin{eqnarray}
  1 &=& U {1 \over 2N_c} \sum_{\bf k} {1 \over {E^{\pm}_{\bf k}}}\,,\\
  E^{(\pm)}_{\bf k} &=& \sqrt{(\epsilon^{(\pm)}_{\bf k} - \mu)^2 + (\Delta^{(\pm)})^2}\,,
\end{eqnarray}
with the Bogoliubov-Valatin quasiparticles given by the well-known
transformation:
\begin{eqnarray}
  \gamma^{(\pm)}_{{\bf k}\uparrow} &=& u^{(\pm)}_{{\bf k}} \chi^{(\pm)}_{{\bf k} \uparrow}
  + v^{(+)}_{\bf k} \chi^{(\pm)\dagger}_{-{\bf k} \downarrow}, \\
    \gamma^{(\pm)\dagger}_{{\bf k}\downarrow} &=& u^{(\pm)}_{{\bf k}} \chi^{(\pm)\dagger}_{{\bf k}\downarrow}  - v^{(+)}_{\bf k} \chi^{(\pm)}_{-{\bf k} \uparrow},
    \end{eqnarray}
where,
\begin{eqnarray}
  u^{(\pm)}_{\mathbf{k}} &=& \left [{1 \over 2} \left( 1 - {{\epsilon^{(\pm)}_\mathbf{k}-\mu}\over {E^{(\pm)}_{\bf k}}} \right ) \right ]^{1/2},  \\
  v^{(\pm)}_{\mathbf{k}} &=& \left [{1 \over 2} \left ( 1 + {{\epsilon^{(\pm)}_\mathbf{k}-\mu}\over {E^{(\pm)}_{\bf k}}} \right ) \right ]^{1/2}.
  \end{eqnarray}

\section{Results}
\label{sec:Results}
\begin{figure*}[!ht]
\centering
    \includegraphics[width=0.8\textwidth]{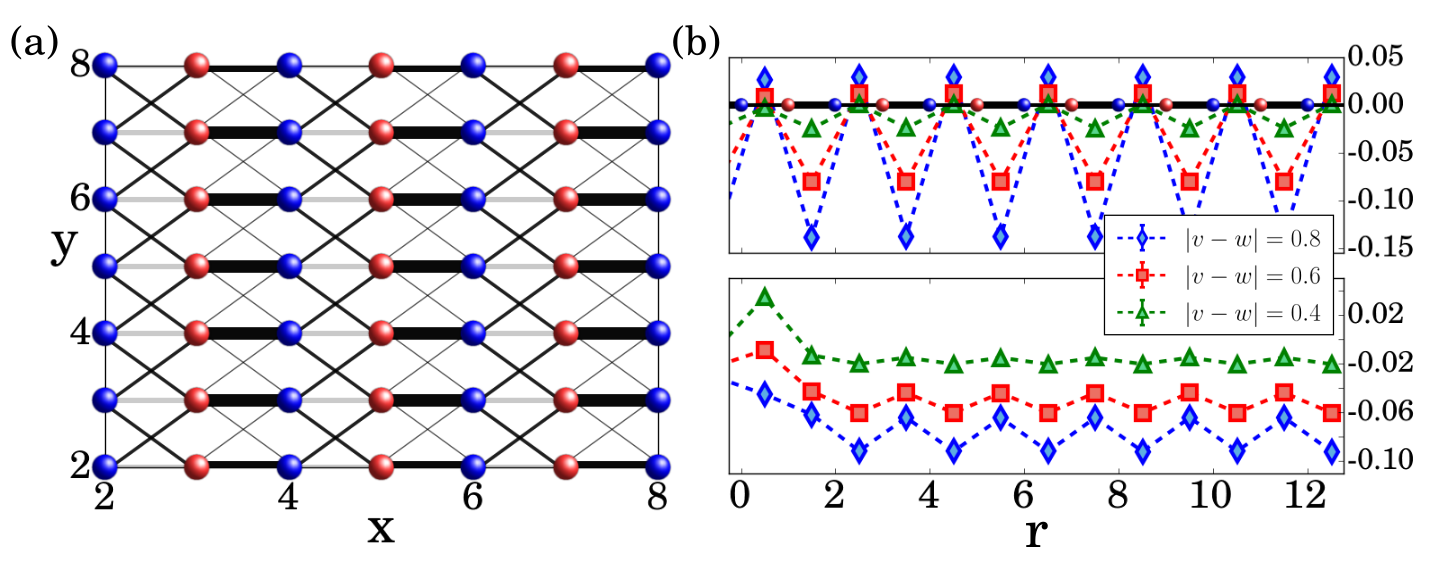}
        \caption{Bond-density order. In (a) we plot a section of $\langle \rho^x_\ell \rho^x_m\rangle$ and $\langle \rho^x_\ell \rho^{\pm x\pm y}_m\rangle$ for $v=0.5, w=1.3$ on a periodic lattice with real-space dimension $26 \times 27$. The upper panel of (b) plots $\langle \rho^x_\ell \rho^x_m\rangle$ for several values of hopping asymmetry, and the lower panel plots $\langle \rho^x_\ell \rho^{\pm x\pm y}_m\rangle$. In the upper panel $\mathbf{r}$
        refers to the $x$-coordinate of sites along $y=6$, and in the lower panel it refers to the $x$-coordinate of sites along the diagonal with origin at $(x,y)=(0,1)$.}
    \label{fig:bond_order}
\end{figure*} 

\begin{figure*}[!ht]
\centering
    \includegraphics[width=0.8\textwidth]{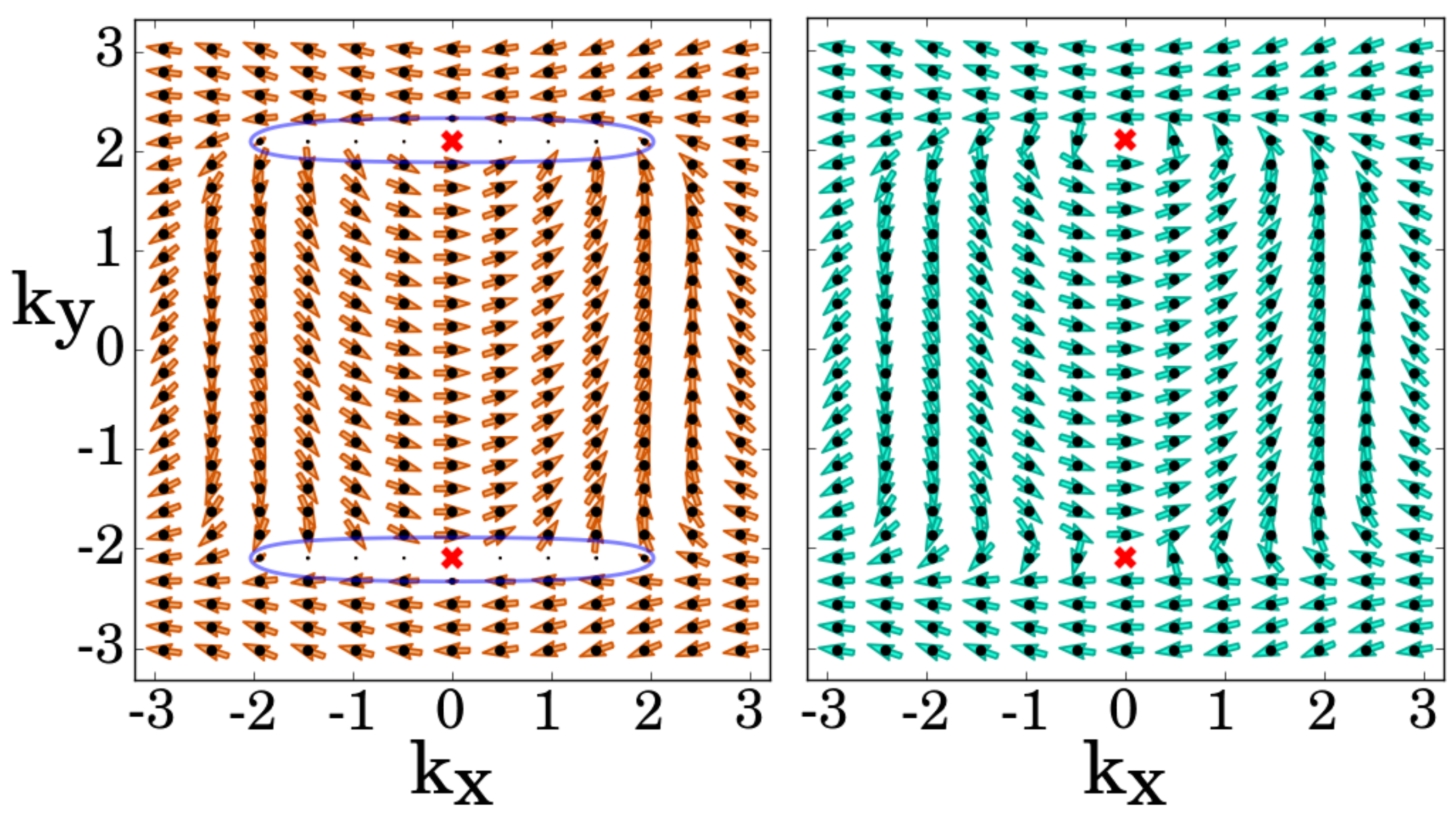}
        \caption{Pseudo-spin  distribution and effective field.  In  the  left  panel  the
          orange  arrows represent  the  many-body expectation  value,
          $(\langle           S^x_\mathbf{k}\rangle,           \langle
          S^y_\mathbf{k}\rangle)$, defined  by Eq.~(\ref{eqn:sk}), which
          gives the  magnitude and  orientation of the  pseudo-spin at
          each lattice momentum.  The dot size is  proportional to the
          total  occupation,   $\langle  n^A_\mathbf{k}+n^B_\mathbf{k}
          \rangle$. The right panel plots the ``effective field''
$\mathbf{h}({\mathbf k})$ given by Eq.~(\ref{hk}). Notice that the          
          pseudo-spin is parallel to the effective field, even in the presence of interactions. The system is a periodic $13\times27$ unit-cell lattice with $v=0.5$, $w=1.3$, $t_d=0.9$, and $U=-1.2$.}
    \label{fig:spin_dist_w_hk}
\end{figure*}

\subsection{Mean-field theory calibration}

As outlined in the preceding section, our approach combines cutting-edge many-body AFQMC calculations with mean-field theory. Our model is free of the fermion sign problem, which means that the AFQMC results are numerically exact and can be used to calibrate the mean-field theory on finite-size systems. We begin this section with a comparison of AFQMC and MFT calculations of the pairing amplitude (defined in Sec.~\ref{sec:pairing_results}) across several parameter sets. This comparison provides a qualitative validation of the mean-field description, which is an important complement to the AFQMC analysis, and also offers an estimate of any finite-size effects in the AFQMC calculations. We find that the mean-field treatment correctly captures the prominent features of the AFQMC results, and that the finite-size effects are generally small.      
We illustrate below the quality of agreement between the MFT and AFQMC results with several representative examples; additional examples are given in Appendix \ref{app-D}.  

In Fig.~\ref{fig:psi_k_mm_QMC_vs_MF} we present a comparison of results for the pairing amplitude in the pseudo-helicity basis from AFQMC and MFT. There is good qualitative agreement between the calculations, which show that pairing occurs primarily in the vicinity of the non-interacting Fermi surface, and that there is relatively little occupation of momentum states above the non-interacting Fermi surface. The MFT result shows a smoother momentum distribution, with occupation spreading above the non-interacting Fermi surface towards the node at $\mathbf{k}=(0,\pm k_N)$, and a correspondingly smooth pairing amplitude, but the essential features of the AFQMC result are evident in the MFT calculation. As we illustrate in Fig.~\ref{fig:psi_k_mm_QMC_MF_path2_vs_v-w} this qualitative agreement holds for increased interaction strength and different values of the hopping asymmetry, $\vert v - w \vert$. Having established a reasonable level of agreement between the AFQMC and MFT results, we can extend our discussion to systems directly in the thermodynamic limit with a certain degree of confidence. Because of their size, these systems are computationally inaccessible to AFQMC, but they are manageable with MFT. 

\subsection{Pseudo-spin distribution and bond-order}

Several features that emerge from the lattice Hamiltonian inform our investigation of the pairing behavior in this system. 
We first consider the pseudo-spin distribution, presented in the left panel of Fig.~\ref{fig:spin_dist_w_hk}. We plot the pseudo-spin direction, $(\langle S^x_\mathbf{k}\rangle,\langle S^y_\mathbf{k}\rangle)$, calculated from the many-body expectation values of the operators,   
\begin{equation}
{\bf{S}}_{\bf{k}} = \frac{1}{2} \sum_{\alpha,\beta,\sigma} \boldsymbol{\sigma}_{\alpha,\beta}  \, 
c^{\dagger(\alpha)}_{{\bf{k}}\sigma} c^{(\beta)}_{{\bf{k}}\sigma},
\label{eqn:sk}
\end{equation}
where $\boldsymbol{\sigma}=(\sigma_x, \sigma_y, \sigma_z)$ is a vector of the Pauli matrices in pseudo-spin space.
On the right side of Fig.~\ref{fig:spin_dist_w_hk} we show the momentum space ``effective field''
$\bf{h}({\bf k})$ given by Eq.~(\ref{hk}). Notice that the          
pseudo-spin is parallel to the effective field, even in the presence of interactions.
This implies that the Weyl quasiparticles in the system experience this field, and the pairing mechanism should be consistent with this effect; in general,
different forms of superconducting order can compete and the ground state
represents the most energetically favorable pairing state in the presence of this pseudo-spin distribution. We elaborate on this connection between the pairing mechanism and the pseudo-spin distribution later in this section.

We next study the effect of the asymmetric hopping along the $\hat{x}$-direction, which results in an oscillation of the bond density, defined by the operator $\rho^{\mu}_{\ell}= c^{\dagger}_{\mathbf{r}_\ell} c_{\mathbf{r}_\ell+\hat{\mu}}+\textrm{h.c.}$, that measures the density on the $\ell$-th bond along the $\mu$-direction, where the first site of the bond is located at position $\mathbf{r}_\ell$. We compute the bond-density correlation function, $\langle \rho^{\mu}_{\ell} \rho^{\nu}_{m}\rangle$, for several values of hopping asymmetry and interaction strength. A typical example is presented in Fig.~\ref{fig:bond_order}. We observe a clear oscillation along the $\hat{x}$-direction that is proportional to the hopping asymmetry (see upper panel in Fig.~\ref{fig:bond_order}(b)). A similar oscillation, with smaller amplitude, is evident along the diagonal directions (lower panel in Fig.~\ref{fig:bond_order}(b)). The magnitude of this order is relatively insensitive to the strength of the interaction, which suggests that the behavior is a primarily one-body effect, though as we illustrate later, the hopping asymmetry responsible for this bond-density wave is also intricately related to the pairing behavior in real and momentum space.

\subsection{Pairing from a topological band structure}
\label{sec:pairing_results}

We now present a detailed picture of the pairing behavior. Our description focuses on the pairing amplitude, which is the eigenstate corresponding to the leading eigenvalue of the two-body density matrix \cite{Yang1962}. We define the elements of this matrix in the pseudo-helicity basis as:
\begin{equation}
M^{\mu\nu \mu^\prime\nu^\prime}_{\mathbf{kq}}=\langle \Delta^{\dagger\mu\nu}_\mathbf{k} \Delta^{\mu^\prime\nu^\prime}_\mathbf{q}\rangle,
\end{equation}
where,
\begin{equation}
    \Delta^{\dagger\mu\nu}_\mathbf{k}=\frac{1}{\sqrt{2}}\left(\chi^{(\mu)\dagger}_{\mathbf{k}\uparrow}\chi^{(\nu)\dagger}_{\mathbf{-k}\downarrow}-\chi^{(\mu)\dagger}_{\mathbf{k}\downarrow}\chi^{(\nu)\dagger}_{\mathbf{-k}\uparrow}\right),
\end{equation}
and $\mu,\nu=\pm$. We note that with an attractive on-site interaction, in the absence of spin-orbit coupling, there is no spin-triplet pairing. Additionally, pairing between pseudo-helicity quasiparticles from different branches ($\mu\neq\nu$ above) is identically zero, so the pairing matrix can be written in $2 N_c\times 2 N_c$ form. 
In this basis the pairing amplitude is,
\begin{equation}
\Psi^p_\mathbf{k} = \psi^{+{}+}_\mathbf{k}+\psi^{-{}-}_\mathbf{k},
\end{equation}
where $\psi^{+{}+}_\mathbf{k}$ ($\psi^{-{}-}_\mathbf{k}$) corresponds to pairs formed from quasiparticles in the upper (lower) pseudo-helicity branch.

In the pseudo-spin basis the pairing amplitude can be written,
\begin{equation}
\Psi^p_\mathbf{k} = \psi^\textmd{AA}_\mathbf{k}+\psi^{\textmd{AB}_t}_\mathbf{k}+\psi^{\textmd{AB}_s}_\mathbf{k}+\psi^{\textmd{BB}}_\mathbf{k},
\end{equation}
which corresponds to the following set of pair creation operators,
\begin{align}
&\Delta^{\dagger\textmd{AA}}_\mathbf{k}=\frac{1}{\sqrt{2}}\left(c^{(\dagger\textmd{A})}_{\mathbf{k}\uparrow}c^{(\dagger\textmd{A})}_{\mathbf{-k}\downarrow}-c^{\dagger(\textmd{A})}_{\mathbf{k}\downarrow}c^{\dagger(\textmd{A})}_{\mathbf{-k}\uparrow}\right);\notag\\
&\Delta^{\dagger\textmd{BB}}_\mathbf{k}=\frac{1}{\sqrt{2}}\left(c^{\dagger(\textmd{B})}_{\mathbf{k}\uparrow}c^{\dagger(\textmd{B})}_{\mathbf{-k}\downarrow}-c^{\dagger(\textmd{B})}_{\mathbf{k}\downarrow}c^{\dagger(\textmd{B})}_{\mathbf{-k}\uparrow}\right);\notag\\
&\Delta^{\dagger\textmd{AB}_t}_\mathbf{k}=\frac{1}{2}\left\{\left(c^{\dagger(\textmd{A})}_{\mathbf{k}\uparrow}c^{\dagger(\textmd{B})}_{\mathbf{-k}\downarrow}-c^{\dagger(\textmd{A})}_{\mathbf{k}\downarrow}c^{\dagger(\textmd{B})}_{\mathbf{-k}\uparrow}\right)\right.\notag\\
&\quad\quad\quad\quad\quad +\left.\left(c^{\dagger(\textmd{B})}_{\mathbf{k}\uparrow}c^{\dagger(\textmd{A})}_{\mathbf{-k}\downarrow}-c^{\dagger(\textmd{B})}_{\mathbf{k}\downarrow}c^{\dagger(\textmd{A})}_{\mathbf{-k}\uparrow}\right)\right\};\notag\\
&\Delta^{\dagger\textmd{AB}_s}_\mathbf{k}=\frac{1}{2}\left\{\left(c^{\dagger(\textmd{A})}_{\mathbf{k}\uparrow}c^{\dagger(\textmd{B})}_{\mathbf{-k}\downarrow}-c^{\dagger(\textmd{A})}_{\mathbf{k}\downarrow}c^{\dagger(\textmd{B})}_{\mathbf{-k}\uparrow}\right)\right.\notag\\
&\quad\quad\quad\quad\quad -\left.\left(c^{\dagger(\textmd{B})}_{\mathbf{k}\uparrow}c^{\dagger(\textmd{A})}_{\mathbf{-k}\downarrow}-c^{\dagger(\textmd{B})}_{\mathbf{k}\downarrow}c^{\dagger(\textmd{A})}_{\mathbf{-k}\uparrow}\right)\right\}.\notag\\
\end{align}
As in the pseudo-helicity basis, all of these operators create spin singlets. The first three create pseudo-spin triplet pairs, and the fourth creates a pseudo-spin singlet.

\begin{figure*}[!ht]
	\centering
	\includegraphics[width=0.9\textwidth]{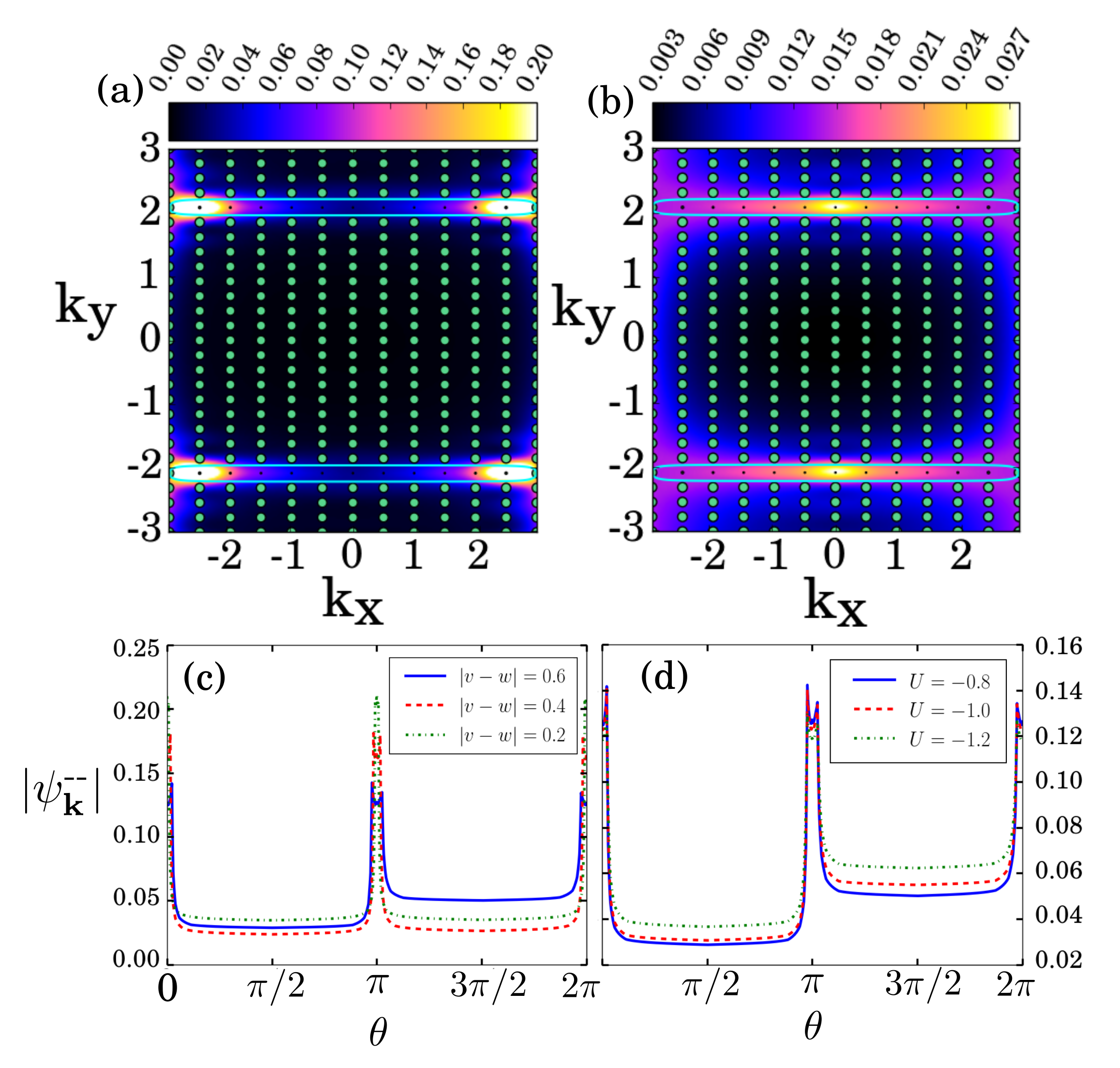}
	\caption{Pairing amplitude in the pseudo-helicity basis from AFQMC. Plotted in (a) is $\vert\psi^{-{}-}_\mathbf{k}\vert$, and in (b) is $\vert\psi^{+{}+}_\mathbf{k}\vert$. The solid light-blue lines represent the non-interacting Fermi surface, and the dot size is  proportional to the total  occupation at a given momentum. The system is a periodic $13 \times 27$ unit-cell lattice with $v=0.7$, $w=1.1$, $t_d=0.9$, and $U=-0.8$. (c) $\vert\psi^{-{}-}_\mathbf{k}\vert$ along the lower Fermi surface versus hopping asymmetry ($t_d=0.9$, and $U=-0.8$). (d) $\vert\psi^{-{}-}_\mathbf{k}\vert$ along the lower Fermi surface versus interaction strength ($v=0.6$, $w=1.2$, $t_d=0.9$). In (c) and (d) $\theta$ is the angle along the Fermi surface (defined using the nodal point, $\mathbf{k}=(0,-k_N)$, as the origin, with $\theta=0$ lying along $k_y=-k_N)$.} 
	\label{fig:psi_k_mm_pp_vs_v-w_vs_U}
\end{figure*}

\begin{figure*}[!ht]
	\centering
	\includegraphics[width=0.9\textwidth]{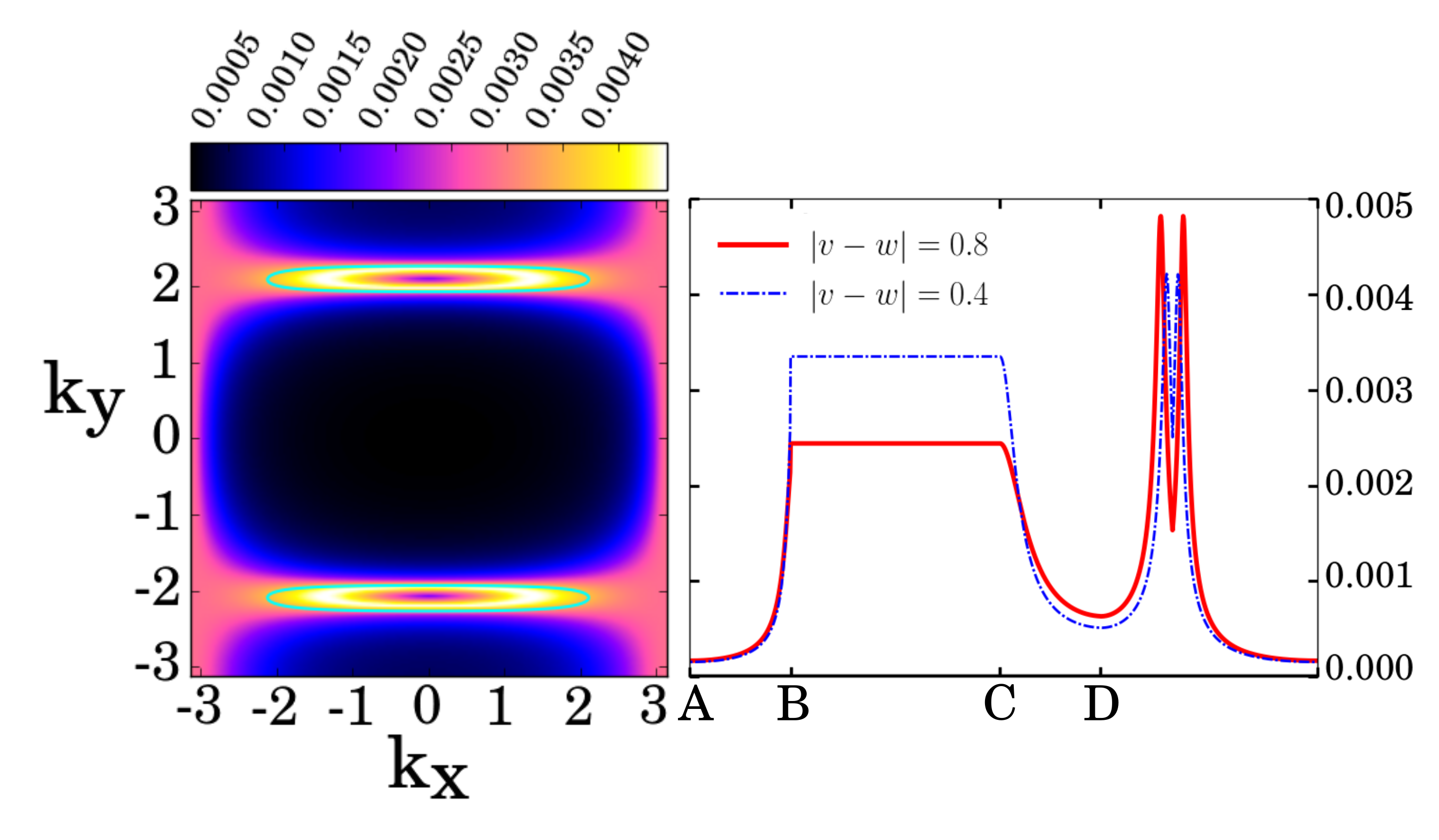}
	\caption{Momentum-space pairing amplitude in the pseudo-helicity basis from MFT. Plotted on the left is $\vert \psi^{-{}-}_\mathbf{k} \vert$ in the thermodynamic limit, with $v=0.6$, $w=1.2$, $t_d=0.9$ and $U=-1.2$. On the right we plot the same quantity along the path drawn in Fig.~\ref{fig:psi_k_mm_QMC_vs_MF}, at different values of hopping asymmetry (with $t_d=0.9$ and $U=-1.2$).}
	\label{fig:psi_k_MF_mm_vs_v-w}
\end{figure*}

\begin{figure*}[!ht]
	\centering
	\includegraphics[width=0.85\textwidth]{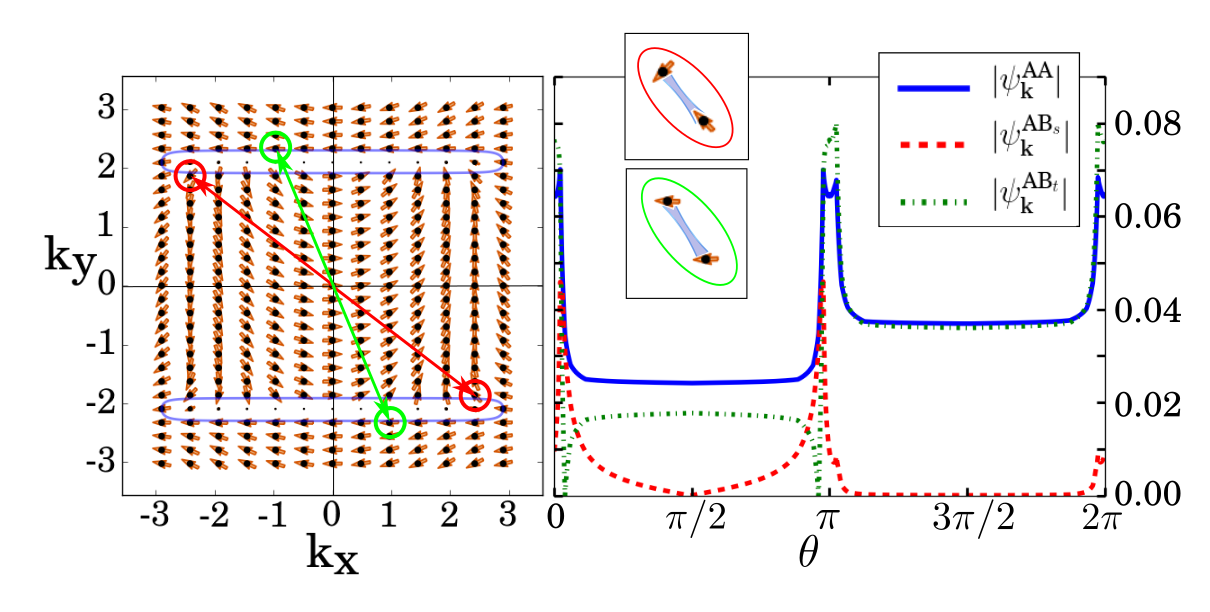}
	\caption{Pseudo-spin distribution and pairing amplitude from AFQMC. In the left panel the orange arrows represent the many-body expectation value, $(\langle S^x_\mathbf{k}\rangle, \langle S^y_\mathbf{k}\rangle)$, defined by Eq.~(\ref{eqn:sk}), which gives the magnitude and orientation of the pseudo-spin at each lattice momentum. The dot size is proportional to the total occupation, $\langle n^\textmd{A}_\mathbf{k}+n^\textmd{B}_\mathbf{k} \rangle$. The right panel plots the various components of the pairing amplitude along the non-interacting Fermi surface (light blue curves in the left panel) centered at $\mathbf{k}=(0,-k_N)$, as in Fig.~\ref{fig:psi_k_mm_pp_vs_v-w_vs_U}. Also illustrated are two examples of the $\mathbf{k}\rightarrow{-\mathbf{k}}$ pairing mechanism that produces pairs with different net pseudo-spin. The system is a periodic $13\times 27$ unit-cell lattice with $v=0.6$, $w=1.2$, $t_d=0.9$, and $U=-1.2$.} 
	\label{fig:spin_dist_w_psi_along_FS}
\end{figure*} 

 \begin{figure*}[!ht]
 	\centering
 	\includegraphics[width=0.95\textwidth]{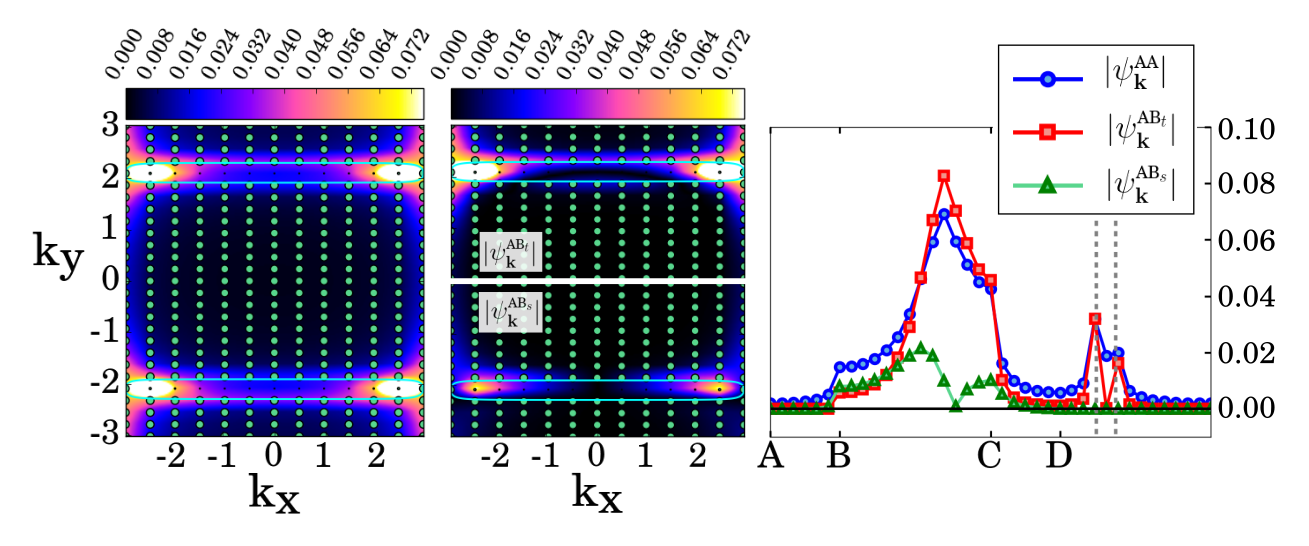}
 	\caption{Momentum-space pairing amplitude in the pseudo-spin basis from AFQMC. Plotted on the left is $\vert\psi^{\textmd{AA}}_\mathbf{k}\vert$. The upper half of the middle panel plots $\vert\psi^{\textmd{AB}_t}_\mathbf{k}\vert$ in the upper half of the BZ, and the lower half plots $\vert\psi^{\textmd{AB}_s}_\mathbf{k}\vert$ in the lower half of the BZ (note that both quantities are symmetric about the $k_x$-axis). On the right, these three components of the pairing amplitude are plotted along the path defined in Fig.~\ref{fig:psi_k_mm_QMC_vs_MF}. Here, $v=0.6$, $w=1.2$, $t_d=0.9$, $U=-0.8$, and the system is a periodic $13\times 27$ unit-cell lattice.}
 	\label{fig:psi_AA_AB_w_loop_path2}
 \end{figure*} 

 \begin{figure*}[!ht]
 	\centering
 	\includegraphics[width=0.95\textwidth]{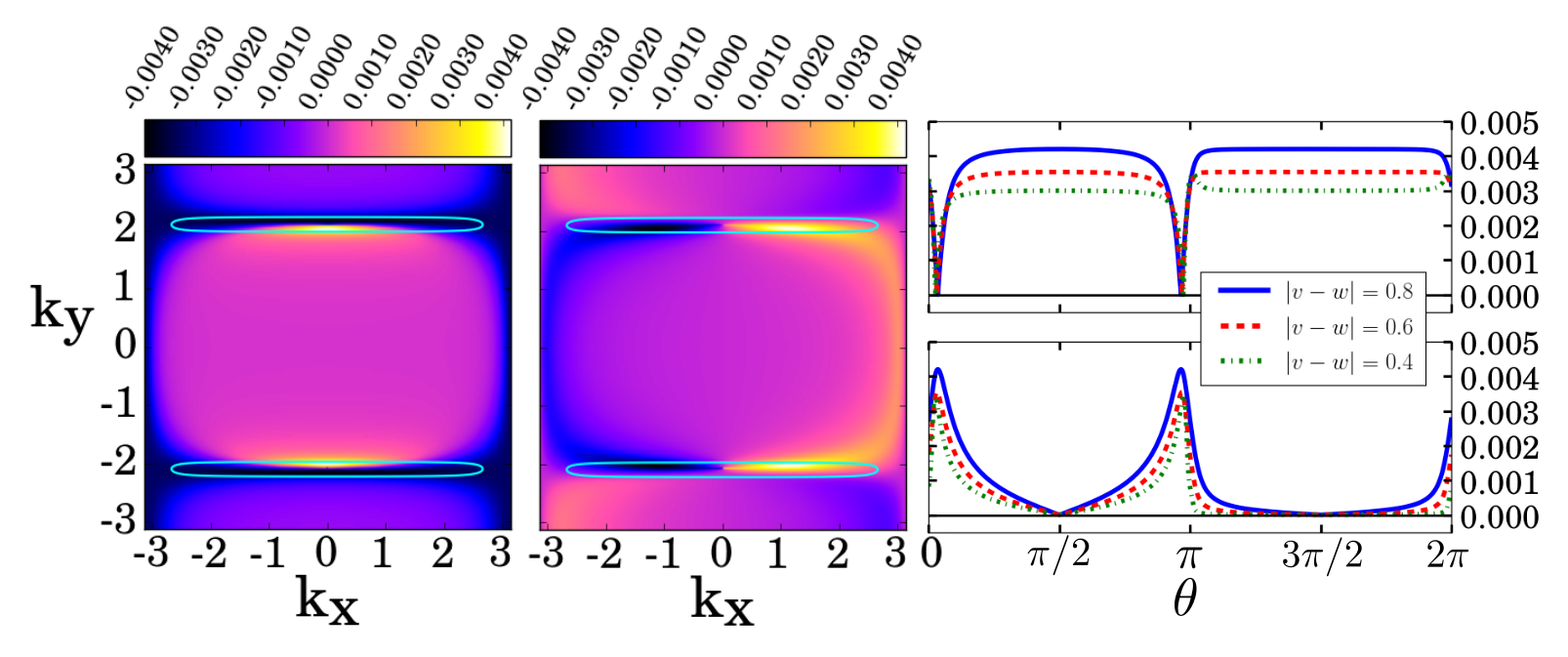}
 	\caption{Momentum-space pairing amplitude in the pseudo-spin basis from MFT. Plotted on the left is the real part of $\psi^{\textmd{AB}_t}_\mathbf{k}$, and in the middle is the imaginary part of $\psi^{\textmd{AB}_s}_\mathbf{k}$ (note that $\psi^{\textmd{AB}_t}_\mathbf{k}$ is purely real, and $\psi^{\textmd{AB}_s}_\mathbf{k}$ is purely imaginary). The system is in the thermodynamic limit, with $v=0.7$, $w=1.1$, $t_d=0.9$ and $U=-1.2$. On the right, the upper panel plots $\vert \psi^{\textmd{AB}_t}_\mathbf{k} \vert$ along the lower Fermi surface (as in Fig.~\ref{fig:psi_k_mm_pp_vs_v-w_vs_U}) for several values of hopping asymmetry (with the remaining parameters equal to those in the left and middle panels), and the lower panel plots $\vert \psi^{\textmd{AB}_s}_\mathbf{k} \vert$ for the same parameters as the upper panel.}
 	\label{fig:psi_MF_AA_AB_w_loop_path2}
 \end{figure*}
 
In the remainder of this section, we analyze in quantitative detail the effects of both interaction and hopping asymmetry on pairing in this system. All the results we present are for densities slightly below half-filling, $n\approx 0.94$, and interaction strengths consistent with the energy scales in relevant materials \cite{MoTe2Rhodes_PRB2017}. The Hamiltonian has particle-hole symmetry, which establishes a direct mapping between electron-doped and hole-doped systems, and implies that the results are symmetric about half-filling.

Figure \ref{fig:psi_k_mm_pp_vs_v-w_vs_U} summarizes the behavior of the pairing amplitude in the pseudo-helicity basis versus interaction strength and hopping asymmetry (see Appendix \ref{app-E} for other parameter values). This quantity is directly related to the gap function, which can be measured via STM or Bogoliubov quasiparticle interference  \cite{STM_Davis2013,STM_Davis2017,Lee2003}. We see significant amplitude for pairing between quasiparticles from the lower helicity band, mostly confined to the vicinity of the non-interacting Fermi surface. Some pairing is evident away from the Fermi surface, which is a reflection of the underlying band structure. The regions where the pairing amplitude has significant magnitude away from the Fermi surface correspond to states near the edges of the Brillouin zone that surround the deep minimum of the lower helicity band. These states are close in energy to the Fermi energy (see Fig.~\ref{fig:band_structure}) and therefore participate in pairing. 
Interestingly, the AFQMC calculations reveal a small amplitude for pairing in the upper helicity band, with a peak near the nodal point (Fig.~\ref{fig:psi_k_mm_pp_vs_v-w_vs_U}(b)). This type of pairing only appears in the mean-field picture for interaction strengths $\vert U\vert \gtrsim 2.0$. As the hopping asymmetry decreases, the curvature of the elliptical Fermi surface decreases, which brings the two sides of the Fermi surface closer together. The effect of this change to the geometry of the Fermi surface on the pairing behavior is evident in the magnitude of the pairing amplitude along the lower Fermi surface (Fig.~\ref{fig:psi_k_mm_pp_vs_v-w_vs_U}(c)). As the hopping asymmetry decreases, there is increased pairing inside the Fermi surface, close to the nodal point, and the pairing is of equal magnitude along either side of the Fermi surface.
The pairing amplitude shows similar quantitative changes with increasing interaction strength, which leads to larger values of the pairing amplitude along the Fermi surface.    
The mean-field description in the thermodynamic limit is consistent with the behavior seen in the AFQMC calculations on finite-size systems (see Fig.~\ref{fig:psi_k_MF_mm_vs_v-w}). Here we find that the pairing amplitude is essentially constant along the non-interacting Fermi surface, and that the pairing amplitude away from the Fermi surface is closely connected to the underlying band structure. As the hopping asymmetry decreases, the helicity bands become flatter and the pairing amplitude away from the Fermi surface increases while the pairing amplitude along the Fermi surface decreases.

Several interesting features of the pairing behavior emerge in the pseudo-spin basis. We explore the connection between the pseudo-spin degree of freedom and pair formation in Fig.~\ref{fig:spin_dist_w_psi_along_FS}. We see from the pseudo-spin distribution that there is a $2\pi$ vortex-like rotation of the pseudo-spin around the upper node at $\mathbf{k}=(0,k_N)$ and a corresponding rotation with opposite vorticity around the lower node at $\mathbf{k}=(0,-k_N)$. This behavior is related to the pairing mechanism, sketched by the circles and connected arrows, which illustrate the formation of zero-momentum pairs along the inner and outer edges of the Fermi surface. Each pair is a real-spin singlet, as well as a mixture of pseudo-spin singlet and pseudo-spin triplet. Pairs formed by electrons in states on the outer edges of the Fermi surface have a large net pseudo-spin triplet component because the electrons in these states have nearly parallel pseudo-spin, and consequently this component of the pairing amplitude has a large magnitude. The same effect diminishes the pairing amplitude in the pseudo-spin singlet sector along the outer edges of the Fermi surface. Along the inner edges, the net pseudo-spin triplet content of the pair is smaller and goes to zero before changing orientation, which leads to a node in the triplet sector of the pairing amplitude. There is also pairing in the pseudo-spin singlet sector along the inner edges of the Fermi surface. This sector of the pairing amplitude exhibits a node at $k_x=0$, corresponding to the momenta at which the pair becomes a net pseudo-spin triplet.    

We present a typical example of the pairing amplitude in the pseudo-spin basis in Fig.~\ref{fig:psi_AA_AB_w_loop_path2}. The left panel plots the AA-sector of the pairing amplitude. This component has significant amplitude along the Fermi surface, and near the edges of the BZ. As in the pseudo-helicity basis, for finite-size systems, the pairing amplitude is larger along the outer edges of the Fermi surface relative to the inner edges. In the middle panel we show the pseudo-spin triplet and singlet sectors, which illustrate the effects of the pseudo-spin degree of freedom in the formation of pairs, as described above. While $\psi^{\textmd{AB}_t}_\mathbf{k}$ has a large amplitude along the outer edges of the Fermi surface, $\psi^{\textmd{AB}_s}_\mathbf{k}$ is zero, whereas along the inner edges, both components are non-zero, with $\psi^{\textmd{AB}_s}_\mathbf{k}$ having a node at $k_x=0$. The right panel, which shows each sector of the pairing amplitude along the path in momentum space drawn in Fig.~\ref{fig:psi_k_mm_QMC_vs_MF}, provides another illustration of the nodal structure of the different components, in particular, the node in $\psi^{\textmd{AB}_t}_\mathbf{k}$ between the inner and outer edges of the Fermi surface.  

\begin{figure*}[!ht]
	\centering
	\includegraphics[width=0.9\textwidth]{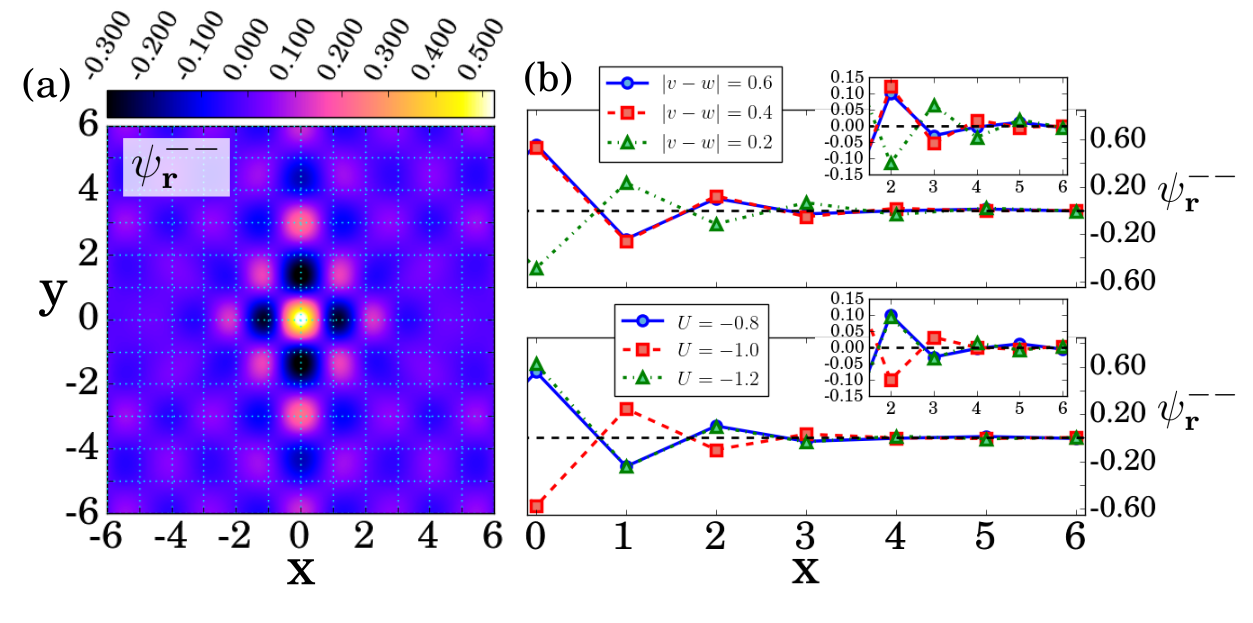}
	\caption{Real space pairing amplitude from AFQMC in the pseudo-helicity basis. In (a) we plot $\psi^{-{}-}_\mathbf{r}$ in the region near the origin, for $v=0.6$, $w=1.2$, $t_d=0.9$, and $U=-0.8$, where $\mathbf{r}$ labels the two-site unit cell. In the upper panel of (b) we show $ \psi^{-{}-}_\mathbf{r}$ along the $x$-axis for several values of $\vert v-w\vert$, with $t_d=0.9$ and $U=-0.8$, and in the lower panel we show the same quantity versus interaction strength, with $v=0.6$, $w=1.2$, $t_d=0.9$. The system is a periodic $13\times 27$ unit-cell lattice.}
	\label{fig:psi_r}
\end{figure*}

\begin{figure*}[!ht]
	\centering
	\includegraphics[width=0.9\textwidth]{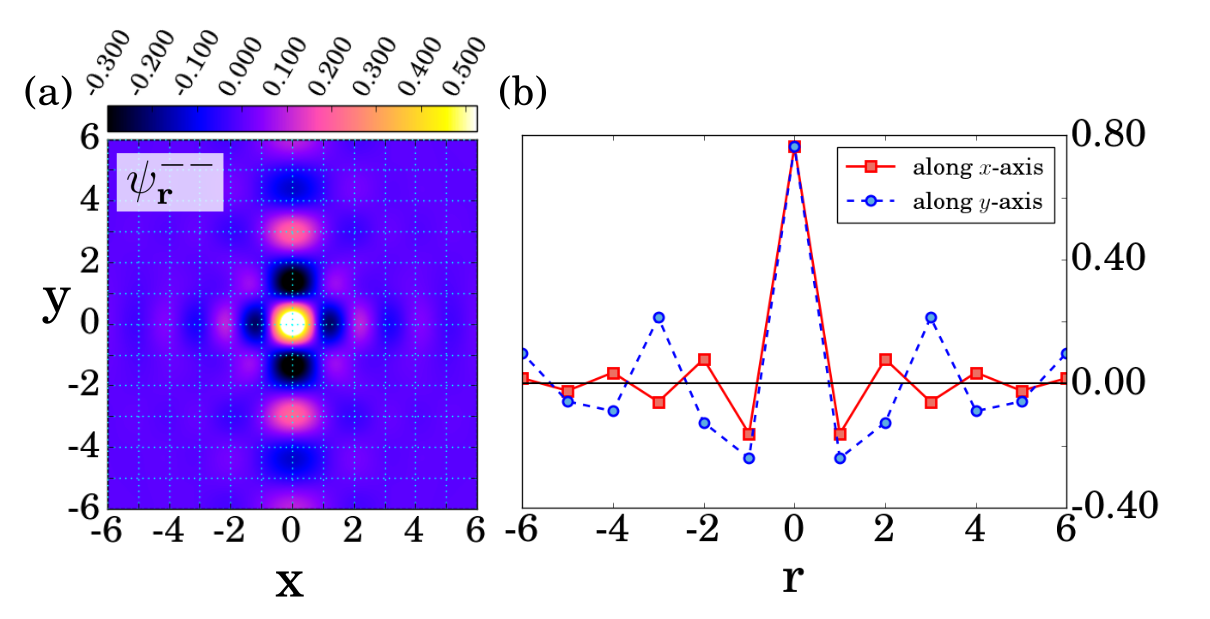}
	\caption{Real space pairing amplitude from MFT in the pseudo-helicity basis. In (a) we plot $\psi^{-{}-}_\mathbf{r}$ from MFT in the thermodynamic limit for the same hopping parameters and interaction strength as Fig.~\ref{fig:psi_r}(a). In (b) we plot slices of $\psi^{-{}-}_\mathbf{r}$ along the $x$-axis and the $y$-axis.}
	\label{fig:psi_r_MF}
\end{figure*}

This exotic pairing behavior also emerges from MFT calculations in the thermodynamic limit, an example of which we present in Fig.~\ref{fig:psi_MF_AA_AB_w_loop_path2}. We see a clear display of the nodal structures of both the pseudo-spin triplet and singlet sectors  of the pairing amplitude. The triplet component shows a sign change along the inner edges of the Fermi surface (left panel of Fig.~\ref{fig:psi_MF_AA_AB_w_loop_path2}), and the singlet component shows a nodal line, dividing the BZ along the $k_y$-axis (middle panel). These nodal structures appear to be largely insensitive to the magnitude of the hopping asymmetry (provided the asymmetry is non-zero).

We find similarly rich pairing properties in real-space.  
Figure \ref{fig:psi_r} presents the real-space structure of the pseudo-helicity pairing amplitude, given by the Fourier transform of $\psi^{-{}-}_\mathbf{k}$. There is a prominent peak at the origin, reflecting the large on-site component of the pair. Away from the origin there are a set of smaller peaks separated by nodes along the $x$ and $y$ axes. We observe that the pairing amplitude becomes more spatially extended as the hopping asymmetry decreases. This behavior is evident in the upper panel of Fig.~\ref{fig:psi_r}(b) and the inset, which shows $\psi^{-{}-}_\mathbf{r}$ along the $x$-axis at decreasing values of $\vert v - w \vert$. The central peak is reduced for smaller values of the hopping asymmetry, while the additional peaks along the axis grow. The lower panel illustrates that the central peak of the pairing amplitude increases with interaction strength, but otherwise the qualitative behavior does not change over the range of interaction strengths we have considered. These behaviors are also evident in the MFT description in the thermodynamic limit (Fig.~\ref{fig:psi_r_MF}).

\section{Discussion and Conclusion}
\label{sec:Discussion_Conclusion}

The discovery of superconducting Weyl materials has ushered in a new era in condensed matter physics. These materials have a number of exciting potential applications in quantum computing and quantum information that have motivated an intense effort to understand their properties.
An accurate quantitative description of these systems at the many-body level remains an essential goal. Such a description presents a unique challenge, demanding a unified treatment of topology and strong correlations. In this work we provide a detailed, high-accuracy characterization of the rich pairing behaviors of a strongly-correlated topological system. We find that both the spin and
pseudo-spin (sublattice) degrees of freedom play an important role, leading to a pairing amplitude with multiple components that have different spatial behaviors. These behaviors, as well as the pairing mechanism, are connected to the pseudo-spin distribution and bond-density order. We observe that in the pseudo-helicity basis pairing occurs only between quasiparticles from the same
pseudo-helicity band carrying opposite topological charge, which results in topological-charge-neutral pairs. 

This topologically neutral form of superconductivity is a consequence of the on-site attractive interaction, which favors the formation of spin-singlet pairs with zero net topological charge.   
The ongoing search for Majorana fermions has indicated that higher order pairing symmetries, for instance $p$-wave pairing, may be an essential ingredient in topological superconductors.
Simple modifications to this model, such as the addition of spin-orbit coupling, or non-local interactions, can induce pairing mechanisms with these different symmetries. As we have demonstrated, the AFQMC method is well-suited to treat these types of topological systems, which makes understanding the origins and mechanisms of topological superconductivity at the many-body level a promising direction for future research.  

Our model offers a simple, fundamental description of the pairing properties that emerge in 
the recently discovered class of superconducting Weyl materials. 
Modern cold atom experiments also offer an ideal platform  
to simulate and study lattice models based on these and other materials. 
There have already been experimental realizations of the Hubbard model \cite{MottInsulatorColdAtoms_Esslinger,Brown1385,Zwierlein_2DFermiHubbard,Parsons1253}, as well as more exotic variations,
including Weyl systems \cite{SSH_coldatoms_2013, PhysRevLett.111.185302,HofstadterModel_AIdelsburger,3DWeylColdAtoms_Ketterle}.
These experiments offer a clean, highly tunable setting in which to explore the intersection of topology and interaction \cite{TopoColdAtoms_RMP}. Our results provide important guidance to this
next generation of experiments exploring the intersection of topological band structures and interaction, and in turn these experiments can serve as a testbed for numerical many-body approaches probing interaction
effects in strongly-correlated topological systems. 

In summary, we have presented an illustrative and quantitative description of the pairing properties of an interacting topological system. Our model contains a pair of Weyl nodes that generate a dipole-like pseudo-spin distribution in momentum space. This distribution exhibits a strong effect on the nature of the pairing amplitude, which represents the most energetically favorable paired state in the presence of the distribution.   
As a consequence, we observe pairing between quasiparticles that carry opposite topological charge, resulting in a topological-charge-neutral pairing amplitude composed of real-spin singlets that are a mixture of pseudo-spin singlet and triplet.  
We provide a thorough characterization of these exotic pairing behaviors using a combination of numerically exact AFQMC calculations and mean-field theory in the thermodynamic limit.
Our results demonstrate the power of the AFQMC method to treat strongly-interacting topologically non-trivial systems at the many-body level. This is an important step towards a more complete understanding of pairing and strong correlations in the context of topological band structures, a subject that has captivated the condensed matter community given the potentially impactful applications of topological superconductivity across quantum computing and quantum information.

\section{Acknowledgments}

This work was supported in part by the U.S. National High Magnetic Field
Laboratory, which is funded by NSF DMR-1157490 and the State of Florida.
Computations were performed in part using the facilities at the College of 
William and Mary which were provided by contributions from the National Science 
Foundation, the Commonwealth of Virginia Equipment Trust Fund and the Office of 
Naval Research.

\appendix
\section{Topological characterization of the non-interacting model: Berry phase and Fermi arcs}
\label{app-A}
\begin{figure}[htb]
	\includegraphics[width=\columnwidth]{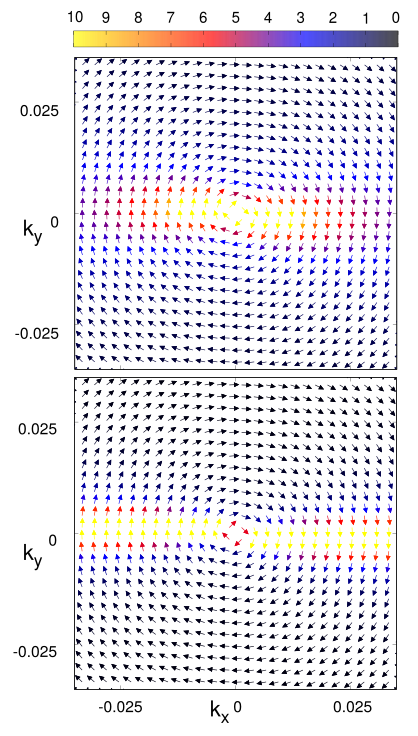}
	\caption{
		Berry potential vector around a single Weyl node. The top panels plots the Berry potential vector (in arbitrary units) as a function of momentum relative to one of the Weyl nodes for the parameters $v=0.6$, $w=1.2$ and $t_d = 0.9$, which yield a fairly isotropic dispersion. The bottom panel plots the same quantity for $v=0.2$, $w=1.0$ and $t_d = 0.4$, which yield a more anisotropic dispersion. The anisotropy of the Berry potential vector is a consequence of the anisotropy of the dispersion.
	}
	\label{fig:vortex}
\end{figure}

In the following section we provide a detailed description of the topological features of our model. As alluded to in the main text, the lattice Hamiltonian given in Eq.~(\ref{hk}) takes the form of a Weyl Hamiltonian with a linear dispersion in the vicinity of the band touching points. These nodal points occur at $\mathbf{k}=(0,\pm k_N)$, with $k_N=\cos^{-1}(-(v+w)/4t_d)$, provided $v+w < 4t_d$. This Weyl form emerges from a Taylor expansion of $\mathbf{h}(\mathbf{k})$ around the nodal points, which yields,
\begin{equation}\mathbf{h}_{\mathrm{linear}}(\mathbf{k})=
\begin{pmatrix}
a_1k_y \\
a_2k_x \\
0
\end{pmatrix},
\label{eqn:hk_linear}
\end{equation}
with 
$a_1=\pm4t_d\mathrm{sin}(k_N)$, and $a_2=(w-v)/2$.  In this case, to leading order in $k_x$ and $k_y$, the dispersion around each node is linear. In the $k_x$-$k_y$ plane, the dispersion is elliptical, with the degree of anisotropy determined by the ratio of $a_1$ to $a_2$.

The Berry potential provides another measure of the topological character of this band structure. In terms of the pseudo-helicity eigenstates, $\vert \pm \rangle$, the Berry potential vector $\mathbf{A}_{\pm}(\mathbf{k}) = \langle \pm | \boldsymbol{\nabla}_{\mathbf{k}} | \pm \rangle$ in the linear regime
is given by:
\begin{equation}
\mathbf{A}_{\pm}(\mathbf{k}) = \frac{a_1a_2}{2a_1^2k_y^2+2a_2^2k_x^2}
\begin{bmatrix}
k_y \\
-k_x \\
0  \\
\end{bmatrix}.
\label{eqn:A_linear}
\end{equation}
As noted in the main text, the Hamiltonian preserves both time-reversal and inversion symmetry, which leads to a vanishing Berry curvature at all points in the BZ, with the exception of the nodal points, where the Berry curvature diverges. Despite this property of the Berry curvature, the Berry potential and the Berry phase around either node are non-zero. Figure~\ref{fig:vortex} plots the vector field $\mathbf{A}_-(\mathbf{k})$ around a single Weyl point for two different values of anisotropy in $a_1$ and $a_2$. The vector field around the other Weyl point has equal magnitude
but opposite vorticity. The top panel of Fig.~\ref{fig:vortex} corresponds to a small anisotropy, while the bottom panel corresponds to a large anisotropy.   
Note that when $a_1 = a_2 $, the vortex pattern of the Berry potential is isotropic.
An isotropic Berry potential and its corresponding dispersion with Weyl nodes, can be engineered by an appropriate choice of hopping parameters, for example, setting $a_1=a_2=1$ gives,
$w = 2 +v$ and $t_d = \sqrt{1+4(1+v)^2}/4$.
The integral of the Berry potential, $\mathbf{A}_\pm(\mathbf{k})$, along any path that encloses a single node gives the Berry phase, which, for the two Weyl nodes in this model is equal to $\pm \pi$. 

For systems with $v+w \geq 4t_d$, the two nodes merge at $\mathbf{k}=(0,\pm \pi)$ and annihilate, which opens a gap in the spectrum without breaking any symmetries of the Hamiltonian.
When the equality above holds (i.e. $v+w = 4t_d$) there is a single nodal point, 
however the dispersion around this point is quadratic.  
In this situation the low energy Hamiltonian is given by:
\begin{equation}\mathbf{h}_{\mathrm{quadratic}}(\mathbf{k})=
\begin{pmatrix}
-b_1k_x^2 + b_2k_y^2 \\
2b_1k_x \\
0
\end{pmatrix},
\label{eqn:hk_quadratic}
\end{equation}
where, 
$b_1=w-2t_d$, and $b_2=2t_d$.
To leading order in $k_x$ the dispersion in this case is linear, however, unlike Eq.~(\ref{eqn:hk_linear}), which is linear in both $k_x$ and $k_y$, this dispersion is quadratic to leading order in $k_y$.

\begin{figure*}[htb]
	\begin{center}
		\subfigure[]{
			\includegraphics[width=3.2 in]{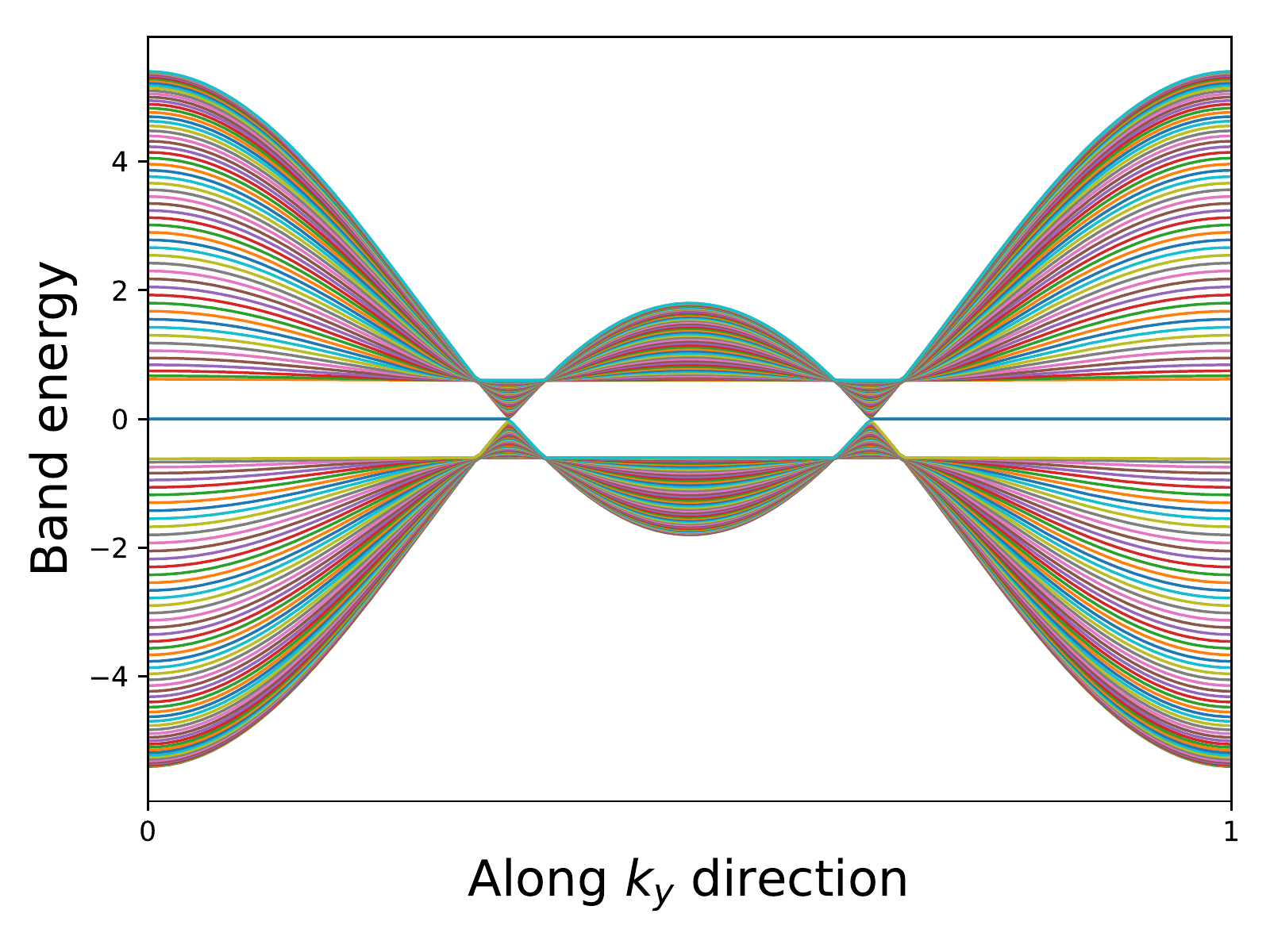}
			\label{fig:along_ky}
		}
		\hskip 0.05 in
		\subfigure[]{
			\includegraphics[width=3.2 in]{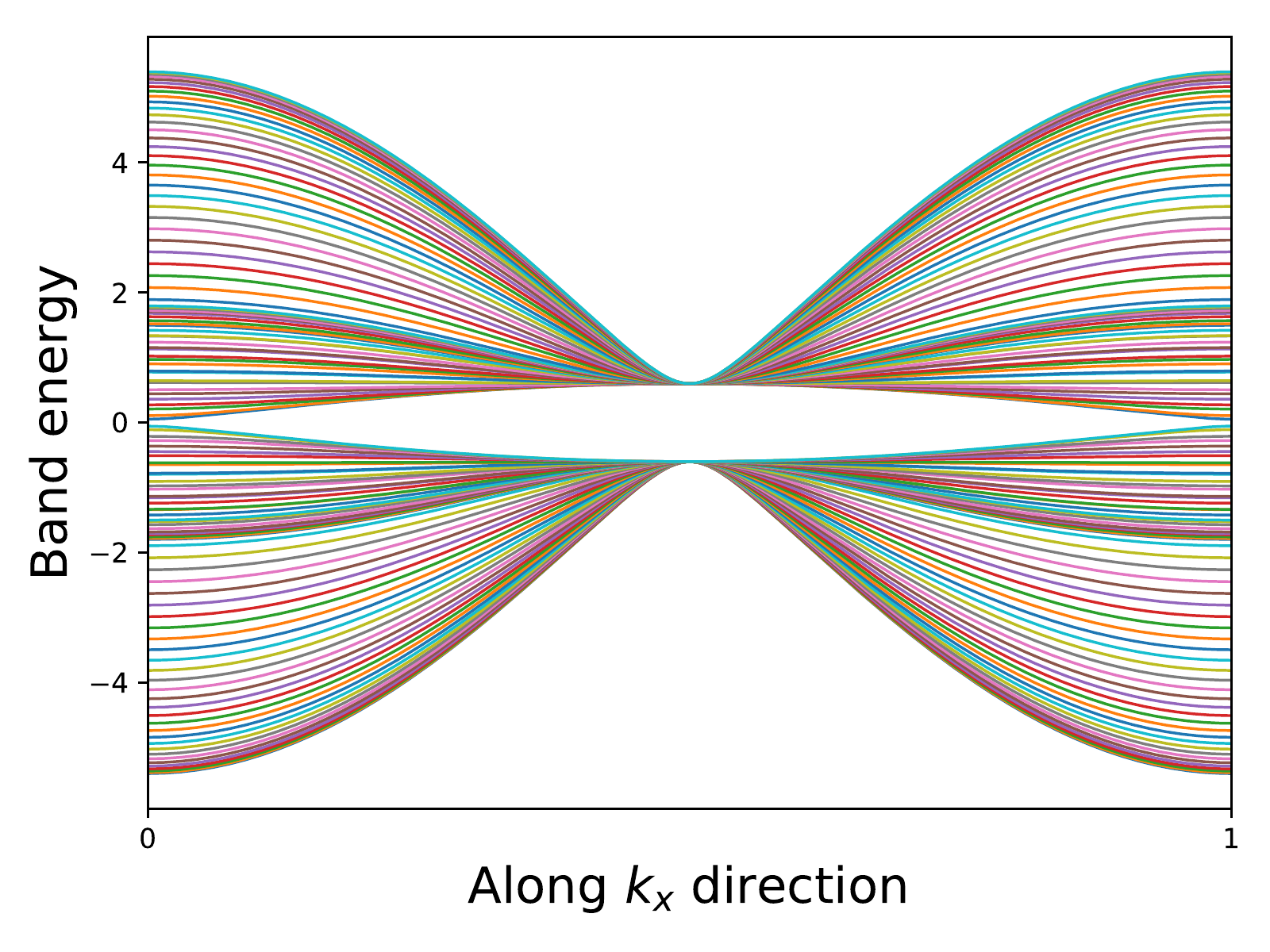}
			\label{fig:along_kx}
		}
		\vskip 0.1 in
		\subfigure[]{
			\includegraphics[width=3.2 in]{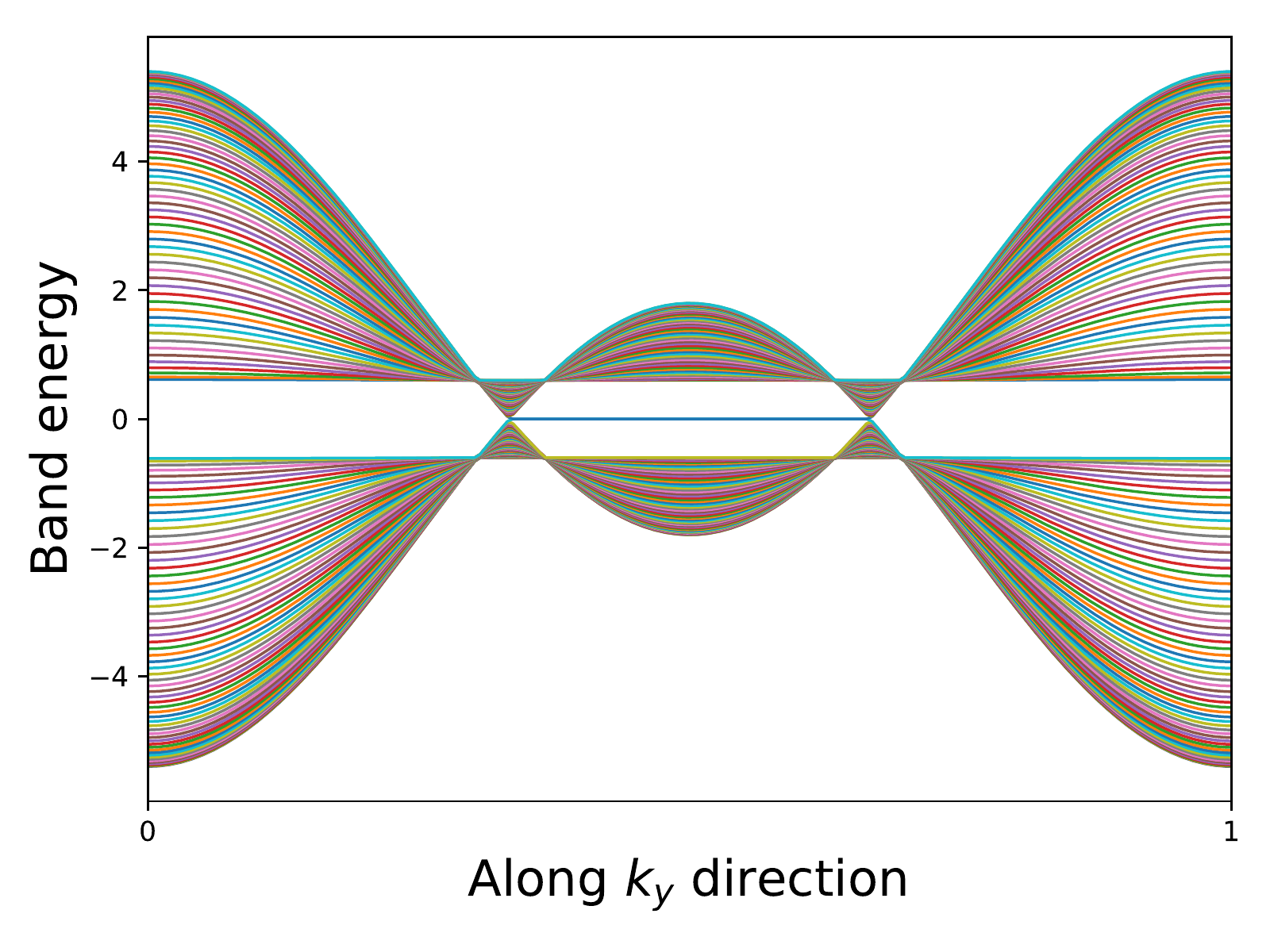}
			\label{fig:along_ky_v_terminated}
		}
	\end{center}
	\caption{
		Band dispersion along (a and c) $k_y$- and (b) $k_x$-direction in units of ${2\pi}/{b}$ and ${2\pi}/{a}$ respectively. In (a) and (c) the lattice has open boundary conditions along $\hat{x}$, and in (b) along $\hat{y}$. The system has parameters $v=0.6$, $w=1.2$ and $t_d = 0.9$.  The dispersion in (c) is the same as that in (a) except that the lattice is terminated on a different bond along the $\hat{x}$-direction. Note how the edge states switch sides depending on the strength of the terminating bond.
	}
	\label{fig:band_dispersion}
\end{figure*}

Finally, we consider the surface spectrum and Fermi arc states.
In Fig.~\ref{fig:band_dispersion} we present the surface spectrum, calculated for a system with open boundary conditions along either the $\hat{x}$- or $\hat{y}$-direction. Figure~\ref{fig:along_ky} plots the dispersion along the $k_y$-direction for a ribbon of finite length in the $\hat{x}$-direction.
The two Weyl nodes at $k_y = \pm k_N$ are connected by dispersionless one dimensional modes, the Fermi arc states,  
which are localized on the two edges of the ribbon.
Such one dimensional edge modes are inherited from the zero dimensional edge states of the SSH chain, which is the building block of our lattice.
These edge states are absent for a ribbon terminated along the $\hat{y}$-direction, as seen in Fig.~\ref{fig:along_kx}.
However, unlike the SSH chain which hosts edge states only when the lattice terminates on the bond with stronger hopping~\cite{asboth2016shortTI}, in our model, edge states are present irrespective of the hopping strength along the terminating bond. The strength of the terminating bond does however determine the orientation of the edge states, as illustrated in Fig.~\ref{fig:along_ky_v_terminated}.

\section{The auxiliary-field quantum Monte Carlo Method}
\label{app-B}
In this section we provide a concise overview of the AFQMC method intended to highlight the essential features of the technique, and refer interested readers to several pedagogical presentations of the formalism for additional details \cite{Lecture-notes,AFQMC_Assaad,Koonin}.

Building on our discussion in the main text, the AFQMC algorithm relies on the projection process defined by Eq.~(\ref{eq:proj_lim}). 
This long imaginary-time projection interval is then divided into
$m = \beta/\delta\tau$ time slices,
\begin{equation}
e^{-\beta\hat{H}}=\left(e^{-\delta\tau\hat{H}}\right)^m,
\end{equation}
which establishes an iterative procedure to obtain the limit in Eq.~(\ref{eq:proj_lim}),
\begin{equation}
\vert\Psi^{(n+1)}\rangle=e^{-\delta\tau\hat{H}}\vert\Psi^{(n)}\rangle,
\label{eq:proj_iter}
\end{equation}
with $\ket{\Psi^{(0)}}=\ket{\Psi_T}$.

Writing our Hamiltonian in the general form, $\hat{H}=\hat{K}+\hat{V}$, where $\hat{K}$ refers to the one-body terms and $\hat{V}$ refers to the two-body terms, we proceed by applying the Trotter-Suzuki decomposition \cite{Trotter,Suzuki}, followed by a Hubbard-Stratonovich transformation \cite{continuous-HS-transformation}:
\begin{equation}
\left(e^{-\delta\tau\hat{H}}\right)^m = 
\left(e^{-\delta\tau\hat{K}/2}e^{-\delta\tau\hat{V}}e^{-\delta\tau\hat{K}/2}\right)^m
+ \mathcal{O}(\delta\tau^2),
\end{equation}
\begin{eqnarray}
e^{-\delta\tau\hat{V}}&=&
e^{-\delta\tau Un^\alpha_{i\uparrow}n^\alpha_{i\downarrow}}\notag\\&=&
\frac{1}{2}\sum_{x^\alpha_i=\pm1}e^{(\gamma x^\alpha_i-\delta\tau U/2)(n^\alpha_{i\uparrow}+n^\alpha_{i\downarrow}-1)},
\label{eq:HS_charge}
\end{eqnarray}
with $\alpha=$~A, B and $\gamma$ is defined by $\cosh(\gamma)=\exp(-\delta\tau\,U/2)$.
Here we have chosen to decouple in the charge channel \cite{HS_transform_discrete}, though decompositions in other channels, such as spin or pairing, exist \cite{AFQMC_RASHBA_INVITE}, and the choice of decomposition can affect the efficiency of the simulation \cite{SymmetryHao}.
This procedure yields the following form for the propagator,
\begin{equation}
e^{-\delta\tau\hat{H}}= \int d\mathbf{x} \,p(\mathbf{x})\hat{B}(\mathbf{x}),
\label{eq:prop_int}
\end{equation}
where $\mathbf{x}=\{x_1,x_2,\hdots,x_{N_S}\}$ is a
set of auxiliary fields at a given time slice, $p(\mathbf{x})$ is a normalized probability density function, and $\hat{B}(\mathbf{x})$ is a one-body operator. The set of auxiliary fields, $\mathbf{x}$, has dimension $N_S$ equal to the
size of the single-particle basis, which in this case is double the number
of unit cells. Note that we have written the propagator in a more general form, as an integral over continuous auxiliary field variables; for the discrete charge decomposition in Eq.~(\ref{eq:HS_charge}) this integral is replaced by a summation over discrete auxiliary field variables. With this choice of Hubbard-Stratonovich transformation $p(\mathbf{x})$ is uniform, and, 
\begin{equation}
\hat{B}(\mathbf{x}) \equiv e^{-\delta\tau\hat{K}/2}
\prod_{i,\alpha} \hat{b}_i(x^\alpha_i)
e^{-\delta\tau\hat{K}/2},
\end{equation}
with $\hat{b}_i(x^\alpha_i) \equiv  \exp\left[{(\gamma x^\alpha_i-\delta\tau U/2)(n^\alpha_{i\uparrow}+n^\alpha_{i\downarrow}-1)}\right]$.
The many-body propagator is now composed of one-body operators
with the fermions in external auxiliary fields. The integration over auxiliary field configurations recovers the two-body interactions.

The many-body, ground-state expectation value of an observable $\hat{O}$ is calculated according to,
\begin{equation}
\langle \hat{O} \rangle = \frac{\bra{\Psi_T}e^{-\beta\hat{H}/2}\hat{O}e^{-\beta\hat{H}/2}\ket{\Psi_T}}
{\bra{\Psi_T}e^{-\beta\hat{H}}\ket{\Psi_T}}.
\label{eq:pi}
\end{equation}
Using the definition of the propagator in Eq.~(\ref{eq:prop_int}), the denominator in Eq.~(\ref{eq:pi}) can be written,
\begin{align}
&\int\bra{\Psi_T} \prod_{\ell=1}^m d\mathbf{x}^{(\ell)}p(\mathbf{x}^{(\ell)})\hat{B}(\mathbf{x}^{(\ell)})\ket{\Psi_T}\notag\\
\equiv&\int\mathcal{W}(\mathbf{X})d\mathbf{X}\,,
\label{eq:pi_denom}
\end{align}
where,
\begin{equation}
\mathcal{W}(\mathbf{X})=\braket{\Psi_l}{\Psi_r}\prod_{\ell=1}^m p(\mathbf{x}^{(\ell)}),
\label{eq:W}
\end{equation}
and we have introduced the notation,
\begin{align*}
\bra{\Psi_l}&=\bra{\Psi_T}\hat{B}(\mathbf{x}^{(m)})\hat{B}(\mathbf{x}^{(m-1)})\hdots\hat{B}(\mathbf{x}^{(n)})\\
\ket{\Psi_r}&=\hat{B}(\mathbf{x}^{(n-1)})\hat{B}(\mathbf{x}^{(n-2)})\hdots\hat{B}(\mathbf{x}^{(1)})\ket{\Psi_T}.
\end{align*}
In the above, $\mathbf{x}^{(\ell)}$ represents an auxiliary field configuration at time slice $\ell$,
and the collection of auxiliary fields $\mathbf{X}=\{\mathbf{x}^{(1)},\mathbf{x}^{(2)},\hdots,\mathbf{x}^{(m)}\}$
defines a path in auxiliary field space.

The expectation value in Eq.~(\ref{eq:pi}) can now be recast as a path integral in auxiliary-field space,
\begin{equation}
\langle \hat{O} \rangle = \frac{\int\mathcal{O}(\mathbf{X})\mathcal{W}(\mathbf{X})d\mathbf{X}}
{\int\mathcal{W}(\mathbf{X})d\mathbf{X}},
\label{eq:pi_shorthand}
\end{equation}
with,
\begin{equation}
\mathcal{O} = \frac{\bra{\Psi_l}\hat{O}\ket{\Psi_r}}{\braket{\Psi_l}{\Psi_r}}.
\end{equation} 
This integral can be evaluated using standard
Monte Carlo techniques, such as the Metropolis algorithm, which
samples auxiliary-fields from $\mathcal{W}(\mathbf{X})$ to obtain a
Monte Carlo estimate of the expectation value in Eq.~(\ref{eq:pi_shorthand}).
To accelerate the sampling procedure we employ a dynamic force
bias \cite{2DFG_AFQMC, Lecture-notes}, which improves the acceptance ratio 
and consequently the efficiency of the algorithm. In addition, we remove the
infinite variance problem using the bridge link method~\cite{inf_var}.

\section{Calibration of AFQMC results against exact diagonalization}
\label{app-C}
In table \ref{table:ED_vs_QMC} we present a comparison of exact diagonalization and AFQMC 
calculations of the total energy.

\begin{table}[htb]
	\begin{tabular}{cccccccccccccccccc}
		\hline\hline
		\multicolumn{1}{c}{$N$} && \multicolumn{1}{c}{$L_x$} && \multicolumn{1}{c}{$L_y$} && \multicolumn{1}{c}{$U$} && \multicolumn{1}{c}{$v$} && \multicolumn{1}{c}{$w$} && \multicolumn{1}{c}{$t_d$} && \multicolumn{1}{c}{$E_\textmd{ED}$} && \multicolumn{1}{c}{$E_\textmd{QMC}$} \\ \hline
		8 && 4 && 3 && -8.0  && 0.5 && 1.3 && 0.6 && -36.64137 && -36.6459(80) \\
		8 && 4 && 3 && -8.0  && 0.2  && 1.0 && 0.3 &&  -34.30740 && -34.3099(76) \\
		2 && 4 && 2 && -12.0  && 0.2 && 0.8 && 0.4 && -12.68698  && -12.6820(43) \\
		4 && 4 && 2 && -12.0  && 0.8  && 1.0 && 0.9 &&  -28.69211 && -28.6930(70) \\
		\hline\hline
	\end{tabular}
	\caption{Comparison of exact diagonalization results with AFQMC results.}
	\label{table:ED_vs_QMC}
\end{table}

\begin{figure}[h]
	\centering
	\includegraphics[width=.85\columnwidth]{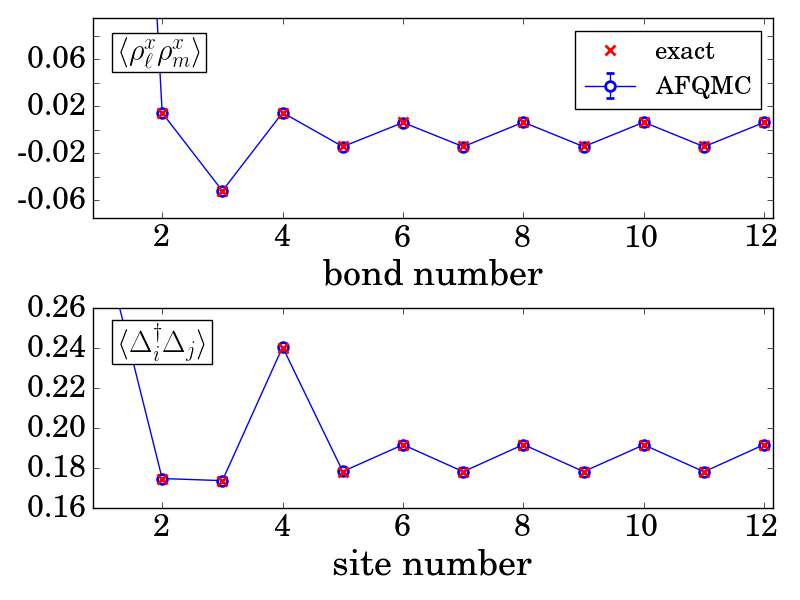}
	\caption{Comparison of exact diagonalization results with AFQMC results for the bond-density and
		s-wave pair-pair correlation functions. The operator $\Delta^\dagger_i = c^\dagger_{i\uparrow}c^\dagger_{i\downarrow}$, where $i$ labels a lattice site. The system is a $2\times 3$ unit-cell lattice with $v=0.6$, $w=1.2$, $t_d=0.9$, $U=-8.0$ and $N=8$ electrons.}
	\label{fig:ED_vs_QMC}
\end{figure}

In Fig.~\ref{fig:ED_vs_QMC} we present a comparison of exact diagonalization and AFQMC 
calculations of different correlation functions.

\section{Additional comparison of AFQMC and mean-field results}
\label{app-D}

\begin{figure}[b]
	\centering
	\includegraphics[width=.825\columnwidth]{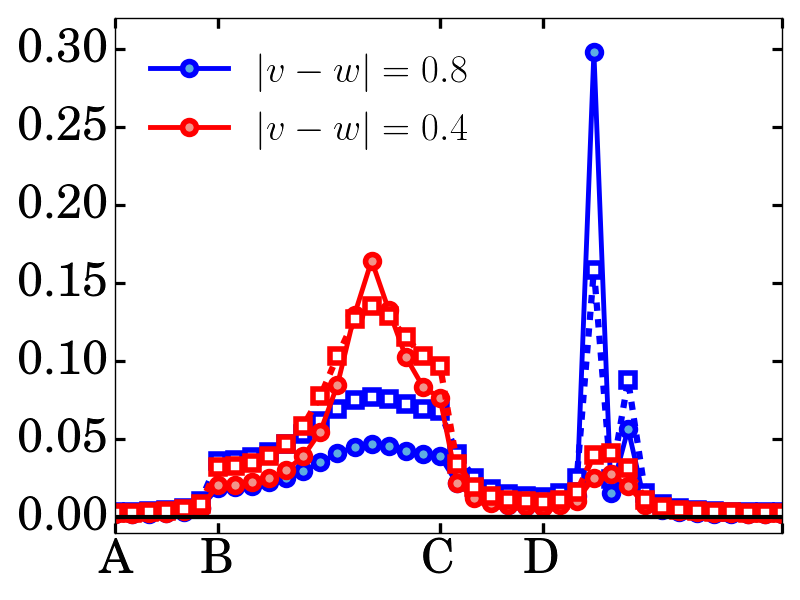}
	\caption{Comparison of momentum-space pairing amplitude in the pseudo-helicity basis from AFQMC and MFT. We show $\vert \psi^{-{}-}_\mathbf{k}\vert$ along the path defined in Fig.~\ref{fig:psi_k_mm_QMC_vs_MF} for two values of hopping asymmetry. The AFQMC result is represented by closed symbols and the MFT result by open symbols. The system is a periodic $13 \times 27$ unit-cell lattice with $t_d=0.9$ and $U=-0.8$}
	\label{fig:QMC_vs_MF_path2}
\end{figure}

\begin{figure}[t]
	\centering
	\includegraphics[width=.8\columnwidth]{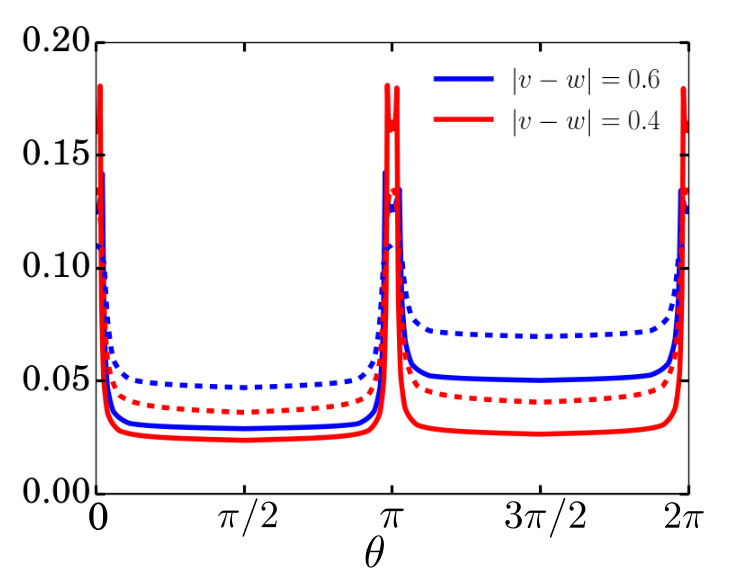}
	\caption{Comparison of momentum-space pairing amplitude in the pseudo-helicity basis along Fermi surface from AFQMC and MFT. We plot $\vert \psi^{-{}-}_\mathbf{k}\vert$ along the non-interacting Fermi surface, with $\theta$ defined as in Fig.~\ref{fig:psi_k_mm_pp_vs_v-w_vs_U}. The solid lines represent the AFQMC result and the dashed lines represent the MFT result. The system is a periodic $13 \times 27$ unit-cell lattice with $t_d=0.9$ and $U=-0.8$.}
	\label{fig:QMC_vs_MF_along_FS}
\end{figure}

In this section we provide results to supplement the comparisons between AFQMC and MFT calculations shown in the main text. Figure~\ref{fig:QMC_vs_MF_path2} shows a similar level of qualitative agreement, especially in the limit of small hopping asymmetry. Generally, the AFQMC result has larger magnitude near the non-interacting Fermi surface, while the MFT result is larger away from the Fermi surface. We see similar qualitative agreement in Fig.~\ref{fig:QMC_vs_MF_along_FS}, where both methods show a difference in the magnitude of the pairing amplitude along the outer edges of the Fermi surface relative to the inner edges at large values of hopping asymmetry. This difference decreases with decreasing hopping asymmetry.  

\begin{figure}[!h]
	\centering
	\includegraphics[width=0.875\columnwidth]{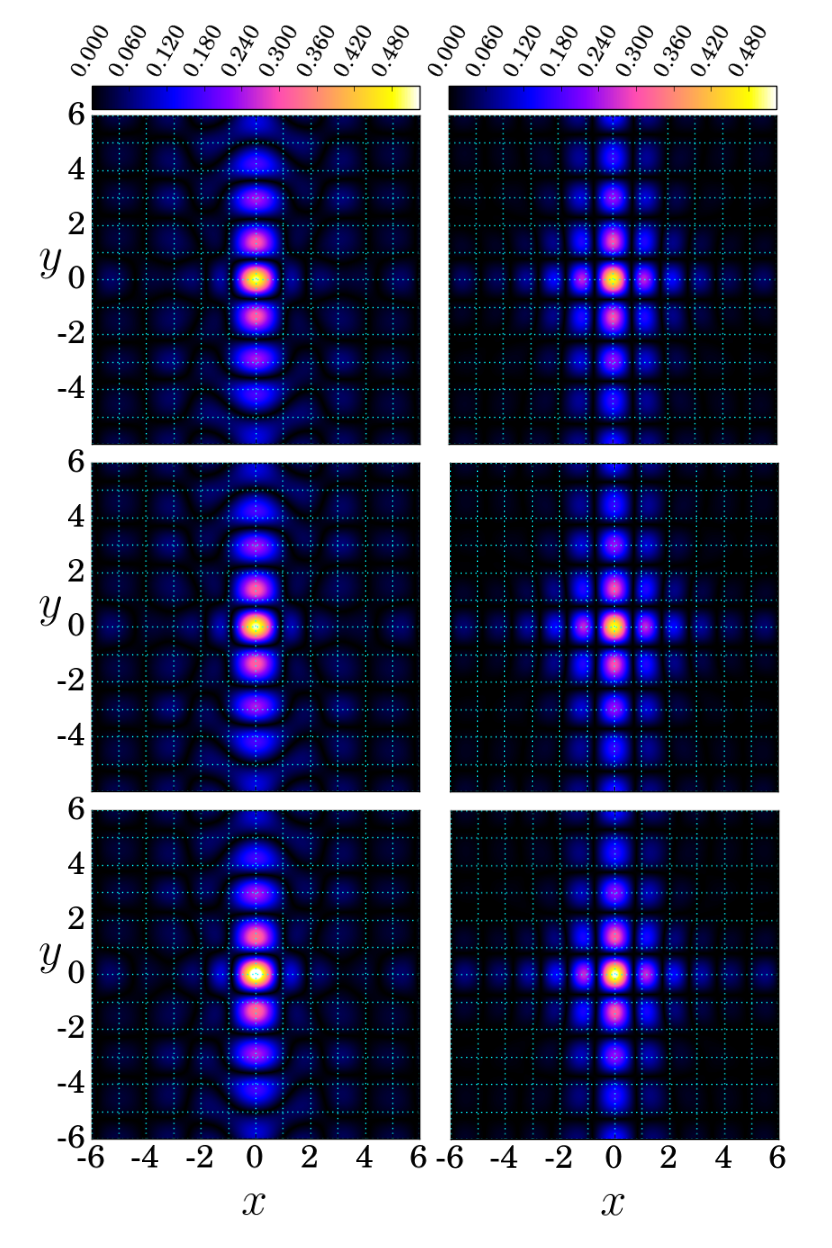}
	\caption{Real-space pairing amplitude in the pseudo-helicity basis. The left column shows $\vert \psi^{-{}-}_\mathbf{r} \vert$ for $v=0.5$, $w=1.3$, $t_d=0.9$, and the right column shows the same quantity for $v=0.8$, $w=1.0$. From top to bottom, $U=-0.8, U=-1.0,U=-1.2$. The system is a periodic $13 \times 27$ unit-cell lattice.}
	\label{fig:psi_r_mm_vs_U_v-w}
\end{figure}

\section{Additional AFQMC results for pairing amplitude}
\label{app-E}

We provide below a survey of results from AFQMC for the pairing amplitude versus interaction strength and hopping asymmetry. In real-space we observe that the central peak of the pairing amplitude grows with interaction strength, but otherwise the qualitative behavior is relatively insensitive to interaction strength over the range of interaction strengths we consider. For smaller values of hopping asymmetry, the pairing amplitude is more spatially extended and isotropic, and still relatively insensitive to the interaction strength.  

\begin{figure*}[!h]
	\centering
	\includegraphics[width=\textwidth]{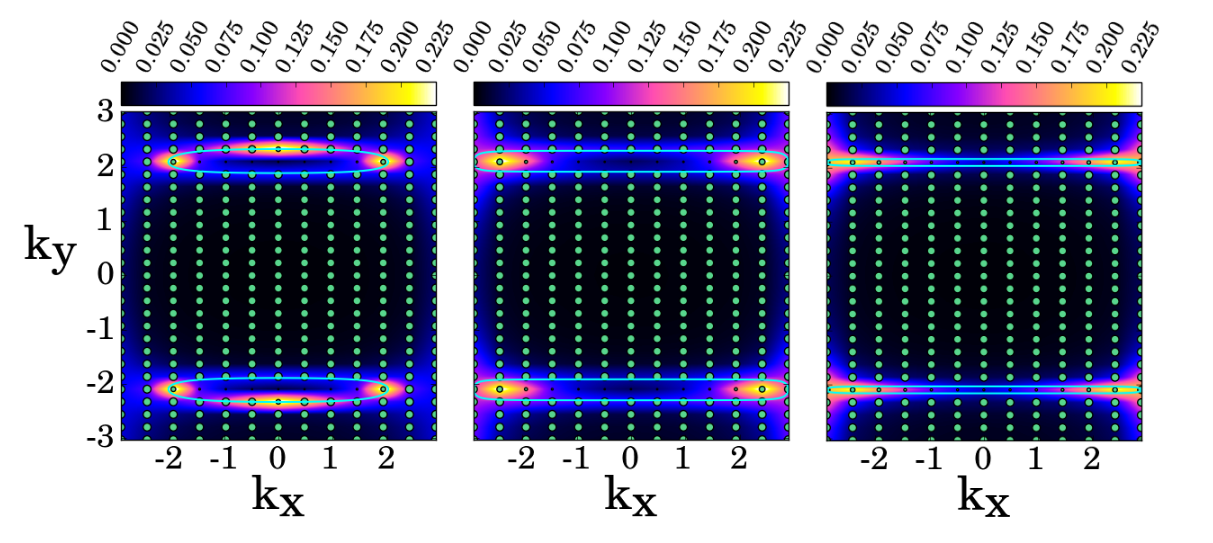}
	\caption{Momentum-space pairing amplitude in the pseudo-helicity basis versus hopping asymmetry. The left panel shows $\vert \psi^{-{}-}_\mathbf{k} \vert$ for $v=0.5$, $w=1.3$, $t_d=0.9$, the middle panel for $v=0.6$, $w=1.2$, and the right panel for $v=0.8$, $w=1.0$. All results are for $t_d=0.9$ and $U=-1.2$, on a periodic $13 \times 27$ unit-cell lattice.}
	\label{fig:psi_k_mm_vs_v-w_U-1.2}
\end{figure*}

%
 

\begin{thebibliography}{48}%
	\makeatletter
	\providecommand \@ifxundefined [1]{%
		\@ifx{#1\undefined}
	}%
	\providecommand \@ifnum [1]{%
		\ifnum #1\expandafter \@firstoftwo
		\else \expandafter \@secondoftwo
		\fi
	}%
	\providecommand \@ifx [1]{%
		\ifx #1\expandafter \@firstoftwo
		\else \expandafter \@secondoftwo
		\fi
	}%
	\providecommand \natexlab [1]{#1}%
	\providecommand \enquote  [1]{``#1''}%
	\providecommand \bibnamefont  [1]{#1}%
	\providecommand \bibfnamefont [1]{#1}%
	\providecommand \citenamefont [1]{#1}%
	\providecommand \href@noop [0]{\@secondoftwo}%
	\providecommand \href [0]{\begingroup \@sanitize@url \@href}%
	\providecommand \@href[1]{\@@startlink{#1}\@@href}%
	\providecommand \@@href[1]{\endgroup#1\@@endlink}%
	\providecommand \@sanitize@url [0]{\catcode `\\12\catcode `\$12\catcode
		`\&12\catcode `\#12\catcode `\^12\catcode `\_12\catcode `\%12\relax}%
	\providecommand \@@startlink[1]{}%
	\providecommand \@@endlink[0]{}%
	\providecommand \url  [0]{\begingroup\@sanitize@url \@url }%
	\providecommand \@url [1]{\endgroup\@href {#1}{\urlprefix }}%
	\providecommand \urlprefix  [0]{URL }%
	\providecommand \Eprint [0]{\href }%
	\providecommand \doibase [0]{http://dx.doi.org/}%
	\providecommand \selectlanguage [0]{\@gobble}%
	\providecommand \bibinfo  [0]{\@secondoftwo}%
	\providecommand \bibfield  [0]{\@secondoftwo}%
	\providecommand \translation [1]{[#1]}%
	\providecommand \BibitemOpen [0]{}%
	\providecommand \bibitemStop [0]{}%
	\providecommand \bibitemNoStop [0]{.\EOS\space}%
	\providecommand \EOS [0]{\spacefactor3000\relax}%
	\providecommand \BibitemShut  [1]{\csname bibitem#1\endcsname}%
	\let\auto@bib@innerbib\@empty
	\bibitem [{\citenamefont {Cheuk}\ \emph
		{et~al.}(2016{\natexlab{a}})\citenamefont {Cheuk}, \citenamefont {Nichols},
		\citenamefont {Lawrence}, \citenamefont {Okan}, \citenamefont {Zhang},
		\citenamefont {Khatami}, \citenamefont {Trivedi}, \citenamefont {Paiva},
		\citenamefont {Rigol},\ and\ \citenamefont {Zwierlein}}]{Cheuk1260}%
	\BibitemOpen
	\bibfield  {author} {\bibinfo {author} {\bibfnamefont {L.~W.}\ \bibnamefont
			{Cheuk}}, \bibinfo {author} {\bibfnamefont {M.~A.}\ \bibnamefont {Nichols}},
		\bibinfo {author} {\bibfnamefont {K.~R.}\ \bibnamefont {Lawrence}}, \bibinfo
		{author} {\bibfnamefont {M.}~\bibnamefont {Okan}}, \bibinfo {author}
		{\bibfnamefont {H.}~\bibnamefont {Zhang}}, \bibinfo {author} {\bibfnamefont
			{E.}~\bibnamefont {Khatami}}, \bibinfo {author} {\bibfnamefont
			{N.}~\bibnamefont {Trivedi}}, \bibinfo {author} {\bibfnamefont
			{T.}~\bibnamefont {Paiva}}, \bibinfo {author} {\bibfnamefont
			{M.}~\bibnamefont {Rigol}}, \ and\ \bibinfo {author} {\bibfnamefont {M.~W.}\
			\bibnamefont {Zwierlein}},\ }\href@noop {} {\bibfield  {journal} {\bibinfo
			{journal} {Science}\ }\textbf {\bibinfo {volume} {353}},\ \bibinfo {pages}
		{1260} (\bibinfo {year} {2016}{\natexlab{a}})}\BibitemShut {NoStop}%
	\bibitem [{\citenamefont {Miyake}\ \emph {et~al.}(2013)\citenamefont {Miyake},
		\citenamefont {Siviloglou}, \citenamefont {Kennedy}, \citenamefont {Burton},\
		and\ \citenamefont {Ketterle}}]{PhysRevLett.111.185302}%
	\BibitemOpen
	\bibfield  {author} {\bibinfo {author} {\bibfnamefont {H.}~\bibnamefont
			{Miyake}}, \bibinfo {author} {\bibfnamefont {G.~A.}\ \bibnamefont
			{Siviloglou}}, \bibinfo {author} {\bibfnamefont {C.~J.}\ \bibnamefont
			{Kennedy}}, \bibinfo {author} {\bibfnamefont {W.~C.}\ \bibnamefont {Burton}},
		\ and\ \bibinfo {author} {\bibfnamefont {W.}~\bibnamefont {Ketterle}},\
	}\href {\doibase 10.1103/PhysRevLett.111.185302} {\bibfield  {journal}
	{\bibinfo  {journal} {Phys. Rev. Lett.}\ }\textbf {\bibinfo {volume} {111}},\
	\bibinfo {pages} {185302} (\bibinfo {year} {2013})}\BibitemShut {NoStop}%
\bibitem [{\citenamefont {Guguchia}\ \emph {et~al.}(2017)\citenamefont
	{Guguchia}, \citenamefont {von Rohr}, \citenamefont {Shermadini},
	\citenamefont {Lee}, \citenamefont {Banerjee}, \citenamefont {Wieteska},
	\citenamefont {Marianetti}, \citenamefont {Frandsen}, \citenamefont
	{Luetkens}, \citenamefont {Gong}, \citenamefont {Cheung}, \citenamefont
	{Baines}, \citenamefont {Shengelaya}, \citenamefont {Taniashvili},
	\citenamefont {Pasupathy}, \citenamefont {Morenzoni}, \citenamefont
	{Billinge}, \citenamefont {Amato}, \citenamefont {Cava}, \citenamefont
	{Khasanov},\ and\ \citenamefont {Uemura}}]{Guguchia2017}%
\BibitemOpen
\bibfield  {author} {\bibinfo {author} {\bibfnamefont {Z.}~\bibnamefont
		{Guguchia}}, \bibinfo {author} {\bibfnamefont {F.}~\bibnamefont {von Rohr}},
	\bibinfo {author} {\bibfnamefont {Z.}~\bibnamefont {Shermadini}}, \bibinfo
	{author} {\bibfnamefont {A.~T.}\ \bibnamefont {Lee}}, \bibinfo {author}
	{\bibfnamefont {S.}~\bibnamefont {Banerjee}}, \bibinfo {author}
	{\bibfnamefont {A.~R.}\ \bibnamefont {Wieteska}}, \bibinfo {author}
	{\bibfnamefont {C.~A.}\ \bibnamefont {Marianetti}}, \bibinfo {author}
	{\bibfnamefont {B.~A.}\ \bibnamefont {Frandsen}}, \bibinfo {author}
	{\bibfnamefont {H.}~\bibnamefont {Luetkens}}, \bibinfo {author}
	{\bibfnamefont {Z.}~\bibnamefont {Gong}}, \bibinfo {author} {\bibfnamefont
		{S.~C.}\ \bibnamefont {Cheung}}, \bibinfo {author} {\bibfnamefont
		{C.}~\bibnamefont {Baines}}, \bibinfo {author} {\bibfnamefont
		{A.}~\bibnamefont {Shengelaya}}, \bibinfo {author} {\bibfnamefont
		{G.}~\bibnamefont {Taniashvili}}, \bibinfo {author} {\bibfnamefont {A.~N.}\
		\bibnamefont {Pasupathy}}, \bibinfo {author} {\bibfnamefont {E.}~\bibnamefont
		{Morenzoni}}, \bibinfo {author} {\bibfnamefont {S.~J.~L.}\ \bibnamefont
		{Billinge}}, \bibinfo {author} {\bibfnamefont {A.}~\bibnamefont {Amato}},
	\bibinfo {author} {\bibfnamefont {R.~J.}\ \bibnamefont {Cava}}, \bibinfo
	{author} {\bibfnamefont {R.}~\bibnamefont {Khasanov}}, \ and\ \bibinfo
	{author} {\bibfnamefont {Y.~J.}\ \bibnamefont {Uemura}},\ }\href {\doibase
	10.1038/s41467-017-01066-6} {\bibfield  {journal} {\bibinfo  {journal}
		{Nature Communications}\ }\textbf {\bibinfo {volume} {8}},\ \bibinfo {pages}
	{1082} (\bibinfo {year} {2017})}\BibitemShut {NoStop}%
\bibitem [{\citenamefont {Rhodes}\ \emph {et~al.}(2017)\citenamefont {Rhodes},
	\citenamefont {Sch\"onemann}, \citenamefont {Aryal}, \citenamefont {Zhou},
	\citenamefont {Zhang}, \citenamefont {Kampert}, \citenamefont {Chiu},
	\citenamefont {Lai}, \citenamefont {Shimura}, \citenamefont {McCandless},
	\citenamefont {Chan}, \citenamefont {Paley}, \citenamefont {Lee},
	\citenamefont {Finke}, \citenamefont {Ruff}, \citenamefont {Das},
	\citenamefont {Manousakis},\ and\ \citenamefont
	{Balicas}}]{MoTe2Rhodes_PRB2017}%
\BibitemOpen
\bibfield  {author} {\bibinfo {author} {\bibfnamefont {D.}~\bibnamefont
		{Rhodes}}, \bibinfo {author} {\bibfnamefont {R.}~\bibnamefont
		{Sch\"onemann}}, \bibinfo {author} {\bibfnamefont {N.}~\bibnamefont {Aryal}},
	\bibinfo {author} {\bibfnamefont {Q.}~\bibnamefont {Zhou}}, \bibinfo {author}
	{\bibfnamefont {Q.~R.}\ \bibnamefont {Zhang}}, \bibinfo {author}
	{\bibfnamefont {E.}~\bibnamefont {Kampert}}, \bibinfo {author} {\bibfnamefont
		{Y.-C.}\ \bibnamefont {Chiu}}, \bibinfo {author} {\bibfnamefont
		{Y.}~\bibnamefont {Lai}}, \bibinfo {author} {\bibfnamefont {Y.}~\bibnamefont
		{Shimura}}, \bibinfo {author} {\bibfnamefont {G.~T.}\ \bibnamefont
		{McCandless}}, \bibinfo {author} {\bibfnamefont {J.~Y.}\ \bibnamefont
		{Chan}}, \bibinfo {author} {\bibfnamefont {D.~W.}\ \bibnamefont {Paley}},
	\bibinfo {author} {\bibfnamefont {J.}~\bibnamefont {Lee}}, \bibinfo {author}
	{\bibfnamefont {A.~D.}\ \bibnamefont {Finke}}, \bibinfo {author}
	{\bibfnamefont {J.~P.~C.}\ \bibnamefont {Ruff}}, \bibinfo {author}
	{\bibfnamefont {S.}~\bibnamefont {Das}}, \bibinfo {author} {\bibfnamefont
		{E.}~\bibnamefont {Manousakis}}, \ and\ \bibinfo {author} {\bibfnamefont
		{L.}~\bibnamefont {Balicas}},\ }\href {\doibase 10.1103/PhysRevB.96.165134}
{\bibfield  {journal} {\bibinfo  {journal} {Phys. Rev. B}\ }\textbf {\bibinfo
		{volume} {96}},\ \bibinfo {pages} {165134} (\bibinfo {year}
	{2017})}\BibitemShut {NoStop}%
\bibitem [{\citenamefont {Rhodes}\ \emph {et~al.}(2019)\citenamefont {Rhodes},
	\citenamefont {Yuan}, \citenamefont {Jung}, \citenamefont {Antony},
	\citenamefont {Wang}, \citenamefont {Kim}, \citenamefont {Chiu},
	\citenamefont {Taniguchi}, \citenamefont {Watanabe}, \citenamefont {Barmak}
	\emph {et~al.}}]{EnhancedTcMoTe2Rhodes_2019}%
\BibitemOpen
\bibfield  {author} {\bibinfo {author} {\bibfnamefont {D.}~\bibnamefont
		{Rhodes}}, \bibinfo {author} {\bibfnamefont {N.~F.}\ \bibnamefont {Yuan}},
	\bibinfo {author} {\bibfnamefont {Y.}~\bibnamefont {Jung}}, \bibinfo {author}
	{\bibfnamefont {A.}~\bibnamefont {Antony}}, \bibinfo {author} {\bibfnamefont
		{H.}~\bibnamefont {Wang}}, \bibinfo {author} {\bibfnamefont {B.}~\bibnamefont
		{Kim}}, \bibinfo {author} {\bibfnamefont {Y.-c.}\ \bibnamefont {Chiu}},
	\bibinfo {author} {\bibfnamefont {T.}~\bibnamefont {Taniguchi}}, \bibinfo
	{author} {\bibfnamefont {K.}~\bibnamefont {Watanabe}}, \bibinfo {author}
	{\bibfnamefont {K.}~\bibnamefont {Barmak}},  \emph {et~al.},\ }\href@noop {}
{\bibfield  {journal} {\bibinfo  {journal} {arXiv preprint arXiv:1905.06508}\
	} (\bibinfo {year} {2019})}\BibitemShut {NoStop}%
\bibitem [{\citenamefont {Yang}\ and\ \citenamefont {Zhang}(2016)}]{Zhang2016}%
\BibitemOpen
\bibfield  {author} {\bibinfo {author} {\bibfnamefont {Z.}~\bibnamefont
		{Yang}}\ and\ \bibinfo {author} {\bibfnamefont {B.}~\bibnamefont {Zhang}},\
}\href {\doibase 10.1103/PhysRevLett.117.224301} {\bibfield  {journal}
{\bibinfo  {journal} {Phys. Rev. Lett.}\ }\textbf {\bibinfo {volume} {117}},\
\bibinfo {pages} {224301} (\bibinfo {year} {2016})}\BibitemShut {NoStop}%
\bibitem [{\citenamefont {Meng}\ and\ \citenamefont
	{Balents}(2012)}]{WeylSuperconductivityBalents}%
\BibitemOpen
\bibfield  {author} {\bibinfo {author} {\bibfnamefont {T.}~\bibnamefont
		{Meng}}\ and\ \bibinfo {author} {\bibfnamefont {L.}~\bibnamefont {Balents}},\
}\href {\doibase 10.1103/PhysRevB.86.054504} {\bibfield  {journal} {\bibinfo
	{journal} {Phys. Rev. B}\ }\textbf {\bibinfo {volume} {86}},\ \bibinfo
{pages} {054504} (\bibinfo {year} {2012})}\BibitemShut {NoStop}%
\bibitem [{\citenamefont {Bednik}\ \emph {et~al.}(2015)\citenamefont {Bednik},
	\citenamefont {Zyuzin},\ and\ \citenamefont
	{Burkov}}]{WeylSuperconductivityBurkov}%
\BibitemOpen
\bibfield  {author} {\bibinfo {author} {\bibfnamefont {G.}~\bibnamefont
		{Bednik}}, \bibinfo {author} {\bibfnamefont {A.~A.}\ \bibnamefont {Zyuzin}},
	\ and\ \bibinfo {author} {\bibfnamefont {A.~A.}\ \bibnamefont {Burkov}},\
}\href {\doibase 10.1103/PhysRevB.92.035153} {\bibfield  {journal} {\bibinfo
	{journal} {Phys. Rev. B}\ }\textbf {\bibinfo {volume} {92}},\ \bibinfo
{pages} {035153} (\bibinfo {year} {2015})}\BibitemShut {NoStop}%
\bibitem [{\citenamefont {Shi}\ \emph {et~al.}(2015{\natexlab{a}})\citenamefont
	{Shi}, \citenamefont {Chiesa},\ and\ \citenamefont {Zhang}}]{2DFG_PRA}%
\BibitemOpen
\bibfield  {author} {\bibinfo {author} {\bibfnamefont {H.}~\bibnamefont
		{Shi}}, \bibinfo {author} {\bibfnamefont {S.}~\bibnamefont {Chiesa}}, \ and\
	\bibinfo {author} {\bibfnamefont {S.}~\bibnamefont {Zhang}},\ }\href
{\doibase 10.1103/PhysRevA.92.033603} {\bibfield  {journal} {\bibinfo
		{journal} {Phys. Rev. A}\ }\textbf {\bibinfo {volume} {92}},\ \bibinfo
	{pages} {033603} (\bibinfo {year} {2015}{\natexlab{a}})}\BibitemShut
{NoStop}%
\bibitem [{\citenamefont {Shi}\ \emph {et~al.}(2016)\citenamefont {Shi},
	\citenamefont {Rosenberg}, \citenamefont {Chiesa},\ and\ \citenamefont
	{Zhang}}]{2DFG_SOC_AFQMC}%
\BibitemOpen
\bibfield  {author} {\bibinfo {author} {\bibfnamefont {H.}~\bibnamefont
		{Shi}}, \bibinfo {author} {\bibfnamefont {P.}~\bibnamefont {Rosenberg}},
	\bibinfo {author} {\bibfnamefont {S.}~\bibnamefont {Chiesa}}, \ and\ \bibinfo
	{author} {\bibfnamefont {S.}~\bibnamefont {Zhang}},\ }\href {\doibase
	10.1103/PhysRevLett.117.040401} {\bibfield  {journal} {\bibinfo  {journal}
		{Phys. Rev. Lett.}\ }\textbf {\bibinfo {volume} {117}},\ \bibinfo {pages}
	{040401} (\bibinfo {year} {2016})}\BibitemShut {NoStop}%
\bibitem [{\citenamefont {Rosenberg}\ \emph {et~al.}(2017)\citenamefont
	{Rosenberg}, \citenamefont {Shi},\ and\ \citenamefont
	{Zhang}}]{AFQMC_Rashba_OPLATT}%
\BibitemOpen
\bibfield  {author} {\bibinfo {author} {\bibfnamefont {P.}~\bibnamefont
		{Rosenberg}}, \bibinfo {author} {\bibfnamefont {H.}~\bibnamefont {Shi}}, \
	and\ \bibinfo {author} {\bibfnamefont {S.}~\bibnamefont {Zhang}},\ }\href
{\doibase 10.1103/PhysRevLett.119.265301} {\bibfield  {journal} {\bibinfo
		{journal} {Phys. Rev. Lett.}\ }\textbf {\bibinfo {volume} {119}},\ \bibinfo
	{pages} {265301} (\bibinfo {year} {2017})}\BibitemShut {NoStop}%
\bibitem [{\citenamefont {Rosenberg}\ \emph {et~al.}(2019)\citenamefont
	{Rosenberg}, \citenamefont {Shi},\ and\ \citenamefont
	{Zhang}}]{AFQMC_RASHBA_INVITE}%
\BibitemOpen
\bibfield  {author} {\bibinfo {author} {\bibfnamefont {P.}~\bibnamefont
		{Rosenberg}}, \bibinfo {author} {\bibfnamefont {H.}~\bibnamefont {Shi}}, \
	and\ \bibinfo {author} {\bibfnamefont {S.}~\bibnamefont {Zhang}},\ }\href
{\doibase https://doi.org/10.1016/j.jpcs.2017.12.026} {\bibfield  {journal}
	{\bibinfo  {journal} {Journal of Physics and Chemistry of Solids}\ }\textbf
	{\bibinfo {volume} {128}},\ \bibinfo {pages} {161} (\bibinfo {year}
	{2019})},\ \bibinfo {note} {{Spin-Orbit Coupled Materials}}\BibitemShut
{NoStop}%
\bibitem [{\citenamefont {Vitali}\ \emph {et~al.}(2016)\citenamefont {Vitali},
	\citenamefont {Shi}, \citenamefont {Qin},\ and\ \citenamefont
	{Zhang}}]{ettoreGAP}%
\BibitemOpen
\bibfield  {author} {\bibinfo {author} {\bibfnamefont {E.}~\bibnamefont
		{Vitali}}, \bibinfo {author} {\bibfnamefont {H.}~\bibnamefont {Shi}},
	\bibinfo {author} {\bibfnamefont {M.}~\bibnamefont {Qin}}, \ and\ \bibinfo
	{author} {\bibfnamefont {S.}~\bibnamefont {Zhang}},\ }\href {\doibase
	10.1103/PhysRevB.94.085140} {\bibfield  {journal} {\bibinfo  {journal} {Phys.
			Rev. B}\ }\textbf {\bibinfo {volume} {94}},\ \bibinfo {pages} {085140}
	(\bibinfo {year} {2016})}\BibitemShut {NoStop}%
\bibitem [{\citenamefont {Otsuka}\ \emph {et~al.}(2018)\citenamefont {Otsuka},
	\citenamefont {Seki}, \citenamefont {Sorella},\ and\ \citenamefont
	{Yunoki}}]{Sorella-2018}%
\BibitemOpen
\bibfield  {author} {\bibinfo {author} {\bibfnamefont {Y.}~\bibnamefont
		{Otsuka}}, \bibinfo {author} {\bibfnamefont {K.}~\bibnamefont {Seki}},
	\bibinfo {author} {\bibfnamefont {S.}~\bibnamefont {Sorella}}, \ and\
	\bibinfo {author} {\bibfnamefont {S.}~\bibnamefont {Yunoki}},\ }\href
{\doibase 10.1103/PhysRevB.98.035126} {\bibfield  {journal} {\bibinfo
		{journal} {Phys. Rev. B}\ }\textbf {\bibinfo {volume} {98}},\ \bibinfo
	{pages} {035126} (\bibinfo {year} {2018})}\BibitemShut {NoStop}%
\bibitem [{\citenamefont {Li}\ \emph {et~al.}(2017)\citenamefont {Li},
	\citenamefont {Jiang},\ and\ \citenamefont {Yao}}]{Yao-2017}%
\BibitemOpen
\bibfield  {author} {\bibinfo {author} {\bibfnamefont {Z.-X.}\ \bibnamefont
		{Li}}, \bibinfo {author} {\bibfnamefont {Y.-F.}\ \bibnamefont {Jiang}}, \
	and\ \bibinfo {author} {\bibfnamefont {H.}~\bibnamefont {Yao}},\ }\href
{\doibase 10.1103/PhysRevLett.119.107202} {\bibfield  {journal} {\bibinfo
		{journal} {Phys. Rev. Lett.}\ }\textbf {\bibinfo {volume} {119}},\ \bibinfo
	{pages} {107202} (\bibinfo {year} {2017})}\BibitemShut {NoStop}%
\bibitem [{\citenamefont {Hung}\ \emph {et~al.}(2014)\citenamefont {Hung},
	\citenamefont {Chua}, \citenamefont {Wang},\ and\ \citenamefont
	{Fiete}}]{Fiete-2014}%
\BibitemOpen
\bibfield  {author} {\bibinfo {author} {\bibfnamefont {H.-H.}\ \bibnamefont
		{Hung}}, \bibinfo {author} {\bibfnamefont {V.}~\bibnamefont {Chua}}, \bibinfo
	{author} {\bibfnamefont {L.}~\bibnamefont {Wang}}, \ and\ \bibinfo {author}
	{\bibfnamefont {G.~A.}\ \bibnamefont {Fiete}},\ }\href {\doibase
	10.1103/PhysRevB.89.235104} {\bibfield  {journal} {\bibinfo  {journal} {Phys.
			Rev. B}\ }\textbf {\bibinfo {volume} {89}},\ \bibinfo {pages} {235104}
	(\bibinfo {year} {2014})}\BibitemShut {NoStop}%
\bibitem [{\citenamefont {Hohenadler}\ \emph {et~al.}(2011)\citenamefont
	{Hohenadler}, \citenamefont {Lang},\ and\ \citenamefont
	{Assaad}}]{Assaad-2011}%
\BibitemOpen
\bibfield  {author} {\bibinfo {author} {\bibfnamefont {M.}~\bibnamefont
		{Hohenadler}}, \bibinfo {author} {\bibfnamefont {T.~C.}\ \bibnamefont
		{Lang}}, \ and\ \bibinfo {author} {\bibfnamefont {F.~F.}\ \bibnamefont
		{Assaad}},\ }\href {\doibase 10.1103/PhysRevLett.106.100403} {\bibfield
	{journal} {\bibinfo  {journal} {Phys. Rev. Lett.}\ }\textbf {\bibinfo
		{volume} {106}},\ \bibinfo {pages} {100403} (\bibinfo {year}
	{2011})}\BibitemShut {NoStop}%
\bibitem [{\citenamefont {Su}\ \emph {et~al.}(1979)\citenamefont {Su},
	\citenamefont {Schrieffer},\ and\ \citenamefont {Heeger}}]{SSH}%
\BibitemOpen
\bibfield  {author} {\bibinfo {author} {\bibfnamefont {W.~P.}\ \bibnamefont
		{Su}}, \bibinfo {author} {\bibfnamefont {J.~R.}\ \bibnamefont {Schrieffer}},
	\ and\ \bibinfo {author} {\bibfnamefont {A.~J.}\ \bibnamefont {Heeger}},\
}\href {\doibase 10.1103/PhysRevLett.42.1698} {\bibfield  {journal} {\bibinfo
	{journal} {Phys. Rev. Lett.}\ }\textbf {\bibinfo {volume} {42}},\ \bibinfo
{pages} {1698} (\bibinfo {year} {1979})}\BibitemShut {NoStop}%
\bibitem [{\citenamefont {Wu}\ and\ \citenamefont
	{Zhang}(2005)}]{nosignproblem}%
\BibitemOpen
\bibfield  {author} {\bibinfo {author} {\bibfnamefont {C.}~\bibnamefont
		{Wu}}\ and\ \bibinfo {author} {\bibfnamefont {S.-C.}\ \bibnamefont {Zhang}},\
}\href {\doibase 10.1103/PhysRevB.71.155115} {\bibfield  {journal} {\bibinfo
	{journal} {Phys. Rev. B}\ }\textbf {\bibinfo {volume} {71}},\ \bibinfo
{pages} {155115} (\bibinfo {year} {2005})}\BibitemShut {NoStop}%
\bibitem [{\citenamefont {Li}\ and\ \citenamefont
	{Haldane}(2018)}]{Haldane-2018}%
\BibitemOpen
\bibfield  {author} {\bibinfo {author} {\bibfnamefont {Y.}~\bibnamefont
		{Li}}\ and\ \bibinfo {author} {\bibfnamefont {F.~D.~M.}\ \bibnamefont
		{Haldane}},\ }\href {\doibase 10.1103/PhysRevLett.120.067003} {\bibfield
	{journal} {\bibinfo  {journal} {Phys. Rev. Lett.}\ }\textbf {\bibinfo
		{volume} {120}},\ \bibinfo {pages} {067003} (\bibinfo {year}
	{2018})}\BibitemShut {NoStop}%
\bibitem [{\citenamefont {Sugiyama}\ and\ \citenamefont
	{Koonin}(1986)}]{Koonin}%
\BibitemOpen
\bibfield  {author} {\bibinfo {author} {\bibfnamefont {G.}~\bibnamefont
		{Sugiyama}}\ and\ \bibinfo {author} {\bibfnamefont {S.}~\bibnamefont
		{Koonin}},\ }\href {\doibase http://dx.doi.org/10.1016/0003-4916(86)90107-7}
{\bibfield  {journal} {\bibinfo  {journal} {Annals of Physics}\ }\textbf
	{\bibinfo {volume} {168}},\ \bibinfo {pages} {1 } (\bibinfo {year}
	{1986})}\BibitemShut {NoStop}%
\bibitem [{\citenamefont {Zhang}(2013)}]{Lecture-notes}%
\BibitemOpen
\bibfield  {author} {\bibinfo {author} {\bibfnamefont {S.}~\bibnamefont
		{Zhang}},\ }in\ \href@noop {} {\emph {\bibinfo {booktitle} {Emergent
			Phenomena in Correlated Matter: Modeling and Simulation}}},\ Vol.~\bibinfo
{volume} {3},\ \bibinfo {editor} {edited by\ \bibinfo {editor} {\bibfnamefont
		{E.}~\bibnamefont {Pavarini}}, \bibinfo {editor} {\bibfnamefont
		{E.}~\bibnamefont {Koch}}, \ and\ \bibinfo {editor} {\bibfnamefont
		{U.}~\bibnamefont {Schollw{\"o}ck}}}\ (\bibinfo  {publisher} {Verlag des
	Forschungszentrum J{\"u}lich, J{\"u}lich, Germany},\ \bibinfo {year}
{2013})\BibitemShut {NoStop}%
\bibitem [{\citenamefont {Assaad}(2002)}]{AFQMC_Assaad}%
\BibitemOpen
\bibfield  {author} {\bibinfo {author} {\bibfnamefont {F.~F.}\ \bibnamefont
		{Assaad}},\ }in\ \href@noop {} {\emph {\bibinfo {booktitle} {Quantum
			Simulations of Complex Many-Body Systems: From Theory to Algorithms}}},\
Vol.~\bibinfo {volume} {10},\ \bibinfo {editor} {edited by\ \bibinfo {editor}
	{\bibfnamefont {J.}~\bibnamefont {Grotendorst}}, \bibinfo {editor}
	{\bibfnamefont {D.}~\bibnamefont {Marx}}, \ and\ \bibinfo {editor}
	{\bibfnamefont {A.}~\bibnamefont {Muramatsu}}}\ (\bibinfo  {publisher} {NIC,
	J{\"u}lich, Germany},\ \bibinfo {year} {2002})\ pp.\ \bibinfo {pages}
{99--156}\BibitemShut {NoStop}%
\bibitem [{\citenamefont {Motta}\ and\ \citenamefont
	{Zhang}(2018)}]{Mario_AFQMC_QC}%
\BibitemOpen
\bibfield  {author} {\bibinfo {author} {\bibfnamefont {M.}~\bibnamefont
		{Motta}}\ and\ \bibinfo {author} {\bibfnamefont {S.}~\bibnamefont {Zhang}},\
}\href {\doibase 10.1002/wcms.1364} {\bibfield  {journal} {\bibinfo
	{journal} {Wiley Interdisciplinary Reviews: Computational Molecular Science}\
}\textbf {\bibinfo {volume} {8}},\ \bibinfo {pages} {e1364} (\bibinfo {year}
{2018})},\ \Eprint
{http://arxiv.org/abs/https://onlinelibrary.wiley.com/doi/pdf/10.1002/wcms.1364}
{https://onlinelibrary.wiley.com/doi/pdf/10.1002/wcms.1364} \BibitemShut
{NoStop}%
\bibitem [{\citenamefont {Landinez~Borda}\ \emph {et~al.}(2019)\citenamefont
	{Landinez~Borda}, \citenamefont {Gomez},\ and\ \citenamefont
	{Morales}}]{Morales_QMC_Multi}%
\BibitemOpen
\bibfield  {author} {\bibinfo {author} {\bibfnamefont {E.~J.}\ \bibnamefont
		{Landinez~Borda}}, \bibinfo {author} {\bibfnamefont {J.}~\bibnamefont
		{Gomez}}, \ and\ \bibinfo {author} {\bibfnamefont {M.~A.}\ \bibnamefont
		{Morales}},\ }\href {\doibase 10.1063/1.5049143} {\bibfield  {journal}
	{\bibinfo  {journal} {The Journal of Chemical Physics}\ }\textbf {\bibinfo
		{volume} {150}},\ \bibinfo {pages} {074105} (\bibinfo {year} {2019})},\
\Eprint {http://arxiv.org/abs/https://doi.org/10.1063/1.5049143}
{https://doi.org/10.1063/1.5049143} \BibitemShut {NoStop}%
\bibitem [{\citenamefont {Qin}\ \emph {et~al.}(2016)\citenamefont {Qin},
	\citenamefont {Shi},\ and\ \citenamefont {Zhang}}]{PhysRevB.94.085103}%
\BibitemOpen
\bibfield  {author} {\bibinfo {author} {\bibfnamefont {M.}~\bibnamefont
		{Qin}}, \bibinfo {author} {\bibfnamefont {H.}~\bibnamefont {Shi}}, \ and\
	\bibinfo {author} {\bibfnamefont {S.}~\bibnamefont {Zhang}},\ }\href
{\doibase 10.1103/PhysRevB.94.085103} {\bibfield  {journal} {\bibinfo
		{journal} {Phys. Rev. B}\ }\textbf {\bibinfo {volume} {94}},\ \bibinfo
	{pages} {085103} (\bibinfo {year} {2016})}\BibitemShut {NoStop}%
\bibitem [{\citenamefont {Vitali}\ \emph {et~al.}(2019)\citenamefont {Vitali},
	\citenamefont {Shi}, \citenamefont {Chiciak},\ and\ \citenamefont
	{Zhang}}]{Ettore_3Band}%
\BibitemOpen
\bibfield  {author} {\bibinfo {author} {\bibfnamefont {E.}~\bibnamefont
		{Vitali}}, \bibinfo {author} {\bibfnamefont {H.}~\bibnamefont {Shi}},
	\bibinfo {author} {\bibfnamefont {A.}~\bibnamefont {Chiciak}}, \ and\
	\bibinfo {author} {\bibfnamefont {S.}~\bibnamefont {Zhang}},\ }\href
{\doibase 10.1103/PhysRevB.99.165116} {\bibfield  {journal} {\bibinfo
		{journal} {Phys. Rev. B}\ }\textbf {\bibinfo {volume} {99}},\ \bibinfo
	{pages} {165116} (\bibinfo {year} {2019})}\BibitemShut {NoStop}%
\bibitem [{\citenamefont {Sun}\ \emph {et~al.}(2013)\citenamefont {Sun},
	\citenamefont {Zhu}, \citenamefont {Liu},\ and\ \citenamefont
	{Ji}}]{2DFG_RASHBASOC_OPLATT}%
\BibitemOpen
\bibfield  {author} {\bibinfo {author} {\bibfnamefont {Q.}~\bibnamefont
		{Sun}}, \bibinfo {author} {\bibfnamefont {G.-B.}\ \bibnamefont {Zhu}},
	\bibinfo {author} {\bibfnamefont {W.-M.}\ \bibnamefont {Liu}}, \ and\
	\bibinfo {author} {\bibfnamefont {A.-C.}\ \bibnamefont {Ji}},\ }\href
{\doibase 10.1103/PhysRevA.88.063637} {\bibfield  {journal} {\bibinfo
		{journal} {Phys. Rev. A}\ }\textbf {\bibinfo {volume} {88}},\ \bibinfo
	{pages} {063637} (\bibinfo {year} {2013})}\BibitemShut {NoStop}%
\bibitem [{\citenamefont {Yang}(1962)}]{Yang1962}%
\BibitemOpen
\bibfield  {author} {\bibinfo {author} {\bibfnamefont {C.~N.}\ \bibnamefont
		{Yang}},\ }\href {\doibase 10.1103/RevModPhys.34.694} {\bibfield  {journal}
	{\bibinfo  {journal} {Rev. Mod. Phys.}\ }\textbf {\bibinfo {volume} {34}},\
	\bibinfo {pages} {694} (\bibinfo {year} {1962})}\BibitemShut {NoStop}%
\bibitem [{\citenamefont {Allan}\ \emph {et~al.}(2013)\citenamefont {Allan},
	\citenamefont {Massee}, \citenamefont {Morr}, \citenamefont {Van~Dyke},
	\citenamefont {Rost}, \citenamefont {Mackenzie}, \citenamefont {Petrovic},\
	and\ \citenamefont {Davis}}]{STM_Davis2013}%
\BibitemOpen
\bibfield  {author} {\bibinfo {author} {\bibfnamefont {M.~P.}\ \bibnamefont
		{Allan}}, \bibinfo {author} {\bibfnamefont {F.}~\bibnamefont {Massee}},
	\bibinfo {author} {\bibfnamefont {D.~K.}\ \bibnamefont {Morr}}, \bibinfo
	{author} {\bibfnamefont {J.}~\bibnamefont {Van~Dyke}}, \bibinfo {author}
	{\bibfnamefont {A.~W.}\ \bibnamefont {Rost}}, \bibinfo {author}
	{\bibfnamefont {A.~P.}\ \bibnamefont {Mackenzie}}, \bibinfo {author}
	{\bibfnamefont {C.}~\bibnamefont {Petrovic}}, \ and\ \bibinfo {author}
	{\bibfnamefont {J.~C.}\ \bibnamefont {Davis}},\ }\href {\doibase
	10.1038/nphys2671} {\bibfield  {journal} {\bibinfo  {journal} {Nature
			Physics}\ }\textbf {\bibinfo {volume} {9}},\ \bibinfo {pages} {468} (\bibinfo
	{year} {2013})}\BibitemShut {NoStop}%
\bibitem [{\citenamefont {Sprau}\ \emph {et~al.}(2017)\citenamefont {Sprau},
	\citenamefont {Kostin}, \citenamefont {Kreisel}, \citenamefont {B{\"o}hmer},
	\citenamefont {Taufour}, \citenamefont {Canfield}, \citenamefont {Mukherjee},
	\citenamefont {Hirschfeld}, \citenamefont {Andersen},\ and\ \citenamefont
	{Davis}}]{STM_Davis2017}%
\BibitemOpen
\bibfield  {author} {\bibinfo {author} {\bibfnamefont {P.~O.}\ \bibnamefont
		{Sprau}}, \bibinfo {author} {\bibfnamefont {A.}~\bibnamefont {Kostin}},
	\bibinfo {author} {\bibfnamefont {A.}~\bibnamefont {Kreisel}}, \bibinfo
	{author} {\bibfnamefont {A.~E.}\ \bibnamefont {B{\"o}hmer}}, \bibinfo
	{author} {\bibfnamefont {V.}~\bibnamefont {Taufour}}, \bibinfo {author}
	{\bibfnamefont {P.~C.}\ \bibnamefont {Canfield}}, \bibinfo {author}
	{\bibfnamefont {S.}~\bibnamefont {Mukherjee}}, \bibinfo {author}
	{\bibfnamefont {P.~J.}\ \bibnamefont {Hirschfeld}}, \bibinfo {author}
	{\bibfnamefont {B.~M.}\ \bibnamefont {Andersen}}, \ and\ \bibinfo {author}
	{\bibfnamefont {J.~C.~S.}\ \bibnamefont {Davis}},\ }\href {\doibase
	10.1126/science.aal1575} {\bibfield  {journal} {\bibinfo  {journal}
		{Science}\ }\textbf {\bibinfo {volume} {357}},\ \bibinfo {pages} {75}
	(\bibinfo {year} {2017})},\ \Eprint
{http://arxiv.org/abs/https://science.sciencemag.org/content/357/6346/75.full.pdf}
{https://science.sciencemag.org/content/357/6346/75.full.pdf} \BibitemShut
{NoStop}%
\bibitem [{\citenamefont {Wang}\ and\ \citenamefont {Lee}(2003)}]{Lee2003}%
\BibitemOpen
\bibfield  {author} {\bibinfo {author} {\bibfnamefont {Q.-H.}\ \bibnamefont
		{Wang}}\ and\ \bibinfo {author} {\bibfnamefont {D.-H.}\ \bibnamefont {Lee}},\
}\href {\doibase 10.1103/PhysRevB.67.020511} {\bibfield  {journal} {\bibinfo
	{journal} {Phys. Rev. B}\ }\textbf {\bibinfo {volume} {67}},\ \bibinfo
{pages} {020511} (\bibinfo {year} {2003})}\BibitemShut {NoStop}%
\bibitem [{\citenamefont {J{\"o}rdens}\ \emph {et~al.}(2008)\citenamefont
	{J{\"o}rdens}, \citenamefont {Strohmaier}, \citenamefont {G{\"u}nter},
	\citenamefont {Moritz},\ and\ \citenamefont
	{Esslinger}}]{MottInsulatorColdAtoms_Esslinger}%
\BibitemOpen
\bibfield  {author} {\bibinfo {author} {\bibfnamefont {R.}~\bibnamefont
		{J{\"o}rdens}}, \bibinfo {author} {\bibfnamefont {N.}~\bibnamefont
		{Strohmaier}}, \bibinfo {author} {\bibfnamefont {K.}~\bibnamefont
		{G{\"u}nter}}, \bibinfo {author} {\bibfnamefont {H.}~\bibnamefont {Moritz}},
	\ and\ \bibinfo {author} {\bibfnamefont {T.}~\bibnamefont {Esslinger}},\
}\href {https://doi.org/10.1038/nature07244} {\bibfield  {journal} {\bibinfo
	{journal} {Nature}\ }\textbf {\bibinfo {volume} {455}},\ \bibinfo {pages}
{204 EP } (\bibinfo {year} {2008})}\BibitemShut {NoStop}%
\bibitem [{\citenamefont {Brown}\ \emph {et~al.}(2017)\citenamefont {Brown},
	\citenamefont {Mitra}, \citenamefont {Guardado-Sanchez}, \citenamefont
	{Schau{\ss}}, \citenamefont {Kondov}, \citenamefont {Khatami}, \citenamefont
	{Paiva}, \citenamefont {Trivedi}, \citenamefont {Huse},\ and\ \citenamefont
	{Bakr}}]{Brown1385}%
\BibitemOpen
\bibfield  {author} {\bibinfo {author} {\bibfnamefont {P.~T.}\ \bibnamefont
		{Brown}}, \bibinfo {author} {\bibfnamefont {D.}~\bibnamefont {Mitra}},
	\bibinfo {author} {\bibfnamefont {E.}~\bibnamefont {Guardado-Sanchez}},
	\bibinfo {author} {\bibfnamefont {P.}~\bibnamefont {Schau{\ss}}}, \bibinfo
	{author} {\bibfnamefont {S.~S.}\ \bibnamefont {Kondov}}, \bibinfo {author}
	{\bibfnamefont {E.}~\bibnamefont {Khatami}}, \bibinfo {author} {\bibfnamefont
		{T.}~\bibnamefont {Paiva}}, \bibinfo {author} {\bibfnamefont
		{N.}~\bibnamefont {Trivedi}}, \bibinfo {author} {\bibfnamefont {D.~A.}\
		\bibnamefont {Huse}}, \ and\ \bibinfo {author} {\bibfnamefont {W.~S.}\
		\bibnamefont {Bakr}},\ }\href {\doibase 10.1126/science.aam7838} {\bibfield
	{journal} {\bibinfo  {journal} {Science}\ }\textbf {\bibinfo {volume}
		{357}},\ \bibinfo {pages} {1385} (\bibinfo {year} {2017})},\ \Eprint
{http://arxiv.org/abs/https://science.sciencemag.org/content/357/6358/1385.full.pdf}
{https://science.sciencemag.org/content/357/6358/1385.full.pdf} \BibitemShut
{NoStop}%
\bibitem [{\citenamefont {Cheuk}\ \emph
	{et~al.}(2016{\natexlab{b}})\citenamefont {Cheuk}, \citenamefont {Nichols},
	\citenamefont {Lawrence}, \citenamefont {Okan}, \citenamefont {Zhang},
	\citenamefont {Khatami}, \citenamefont {Trivedi}, \citenamefont {Paiva},
	\citenamefont {Rigol},\ and\ \citenamefont
	{Zwierlein}}]{Zwierlein_2DFermiHubbard}%
\BibitemOpen
\bibfield  {author} {\bibinfo {author} {\bibfnamefont {L.~W.}\ \bibnamefont
		{Cheuk}}, \bibinfo {author} {\bibfnamefont {M.~A.}\ \bibnamefont {Nichols}},
	\bibinfo {author} {\bibfnamefont {K.~R.}\ \bibnamefont {Lawrence}}, \bibinfo
	{author} {\bibfnamefont {M.}~\bibnamefont {Okan}}, \bibinfo {author}
	{\bibfnamefont {H.}~\bibnamefont {Zhang}}, \bibinfo {author} {\bibfnamefont
		{E.}~\bibnamefont {Khatami}}, \bibinfo {author} {\bibfnamefont
		{N.}~\bibnamefont {Trivedi}}, \bibinfo {author} {\bibfnamefont
		{T.}~\bibnamefont {Paiva}}, \bibinfo {author} {\bibfnamefont
		{M.}~\bibnamefont {Rigol}}, \ and\ \bibinfo {author} {\bibfnamefont {M.~W.}\
		\bibnamefont {Zwierlein}},\ }\href
{http://science.sciencemag.org/content/353/6305/1260.abstract} {\bibfield
	{journal} {\bibinfo  {journal} {Science}\ }\textbf {\bibinfo {volume}
		{353}},\ \bibinfo {pages} {1260} (\bibinfo {year}
	{2016}{\natexlab{b}})}\BibitemShut {NoStop}%
\bibitem [{\citenamefont {Parsons}\ \emph {et~al.}(2016)\citenamefont
	{Parsons}, \citenamefont {Mazurenko}, \citenamefont {Chiu}, \citenamefont
	{Ji}, \citenamefont {Greif},\ and\ \citenamefont {Greiner}}]{Parsons1253}%
\BibitemOpen
\bibfield  {author} {\bibinfo {author} {\bibfnamefont {M.~F.}\ \bibnamefont
		{Parsons}}, \bibinfo {author} {\bibfnamefont {A.}~\bibnamefont {Mazurenko}},
	\bibinfo {author} {\bibfnamefont {C.~S.}\ \bibnamefont {Chiu}}, \bibinfo
	{author} {\bibfnamefont {G.}~\bibnamefont {Ji}}, \bibinfo {author}
	{\bibfnamefont {D.}~\bibnamefont {Greif}}, \ and\ \bibinfo {author}
	{\bibfnamefont {M.}~\bibnamefont {Greiner}},\ }\href@noop {} {\bibfield
	{journal} {\bibinfo  {journal} {Science}\ }\textbf {\bibinfo {volume}
		{353}},\ \bibinfo {pages} {1253} (\bibinfo {year} {2016})}\BibitemShut
{NoStop}%
\bibitem [{\citenamefont {Atala}\ \emph {et~al.}(2013)\citenamefont {Atala},
	\citenamefont {Aidelsburger}, \citenamefont {Barreiro}, \citenamefont
	{Abanin}, \citenamefont {Kitagawa}, \citenamefont {Demler},\ and\
	\citenamefont {Bloch}}]{SSH_coldatoms_2013}%
\BibitemOpen
\bibfield  {author} {\bibinfo {author} {\bibfnamefont {M.}~\bibnamefont
		{Atala}}, \bibinfo {author} {\bibfnamefont {M.}~\bibnamefont {Aidelsburger}},
	\bibinfo {author} {\bibfnamefont {J.~T.}\ \bibnamefont {Barreiro}}, \bibinfo
	{author} {\bibfnamefont {D.}~\bibnamefont {Abanin}}, \bibinfo {author}
	{\bibfnamefont {T.}~\bibnamefont {Kitagawa}}, \bibinfo {author}
	{\bibfnamefont {E.}~\bibnamefont {Demler}}, \ and\ \bibinfo {author}
	{\bibfnamefont {I.}~\bibnamefont {Bloch}},\ }\href
{https://doi.org/10.1038/nphys2790} {\bibfield  {journal} {\bibinfo
		{journal} {Nature Physics}\ }\textbf {\bibinfo {volume} {9}},\ \bibinfo
	{pages} {795} (\bibinfo {year} {2013})},\ \bibinfo {note}
{article}\BibitemShut {NoStop}%
\bibitem [{\citenamefont {Aidelsburger}\ \emph {et~al.}(2013)\citenamefont
	{Aidelsburger}, \citenamefont {Atala}, \citenamefont {Lohse}, \citenamefont
	{Barreiro}, \citenamefont {Paredes},\ and\ \citenamefont
	{Bloch}}]{HofstadterModel_AIdelsburger}%
\BibitemOpen
\bibfield  {author} {\bibinfo {author} {\bibfnamefont {M.}~\bibnamefont
		{Aidelsburger}}, \bibinfo {author} {\bibfnamefont {M.}~\bibnamefont {Atala}},
	\bibinfo {author} {\bibfnamefont {M.}~\bibnamefont {Lohse}}, \bibinfo
	{author} {\bibfnamefont {J.~T.}\ \bibnamefont {Barreiro}}, \bibinfo {author}
	{\bibfnamefont {B.}~\bibnamefont {Paredes}}, \ and\ \bibinfo {author}
	{\bibfnamefont {I.}~\bibnamefont {Bloch}},\ }\href {\doibase
	10.1103/PhysRevLett.111.185301} {\bibfield  {journal} {\bibinfo  {journal}
		{Phys. Rev. Lett.}\ }\textbf {\bibinfo {volume} {111}},\ \bibinfo {pages}
	{185301} (\bibinfo {year} {2013})}\BibitemShut {NoStop}%
\bibitem [{\citenamefont {Dub\ifmmode~\check{c}\else \v{c}\fi{}ek}\ \emph
	{et~al.}(2015)\citenamefont {Dub\ifmmode~\check{c}\else \v{c}\fi{}ek},
	\citenamefont {Kennedy}, \citenamefont {Lu}, \citenamefont {Ketterle},
	\citenamefont {Solja\ifmmode \check{c}\else
		\v{c}\fi{}i\ifmmode~\acute{c}\else \'{c}\fi{}},\ and\ \citenamefont
	{Buljan}}]{3DWeylColdAtoms_Ketterle}%
\BibitemOpen
\bibfield  {author} {\bibinfo {author} {\bibfnamefont {T.}~\bibnamefont
		{Dub\ifmmode~\check{c}\else \v{c}\fi{}ek}}, \bibinfo {author} {\bibfnamefont
		{C.~J.}\ \bibnamefont {Kennedy}}, \bibinfo {author} {\bibfnamefont
		{L.}~\bibnamefont {Lu}}, \bibinfo {author} {\bibfnamefont {W.}~\bibnamefont
		{Ketterle}}, \bibinfo {author} {\bibfnamefont {M.}~\bibnamefont
		{Solja\ifmmode \check{c}\else \v{c}\fi{}i\ifmmode~\acute{c}\else
			\'{c}\fi{}}}, \ and\ \bibinfo {author} {\bibfnamefont {H.}~\bibnamefont
		{Buljan}},\ }\href {\doibase 10.1103/PhysRevLett.114.225301} {\bibfield
	{journal} {\bibinfo  {journal} {Phys. Rev. Lett.}\ }\textbf {\bibinfo
		{volume} {114}},\ \bibinfo {pages} {225301} (\bibinfo {year}
	{2015})}\BibitemShut {NoStop}%
\bibitem [{\citenamefont {Cooper}\ \emph {et~al.}(2019)\citenamefont {Cooper},
	\citenamefont {Dalibard},\ and\ \citenamefont
	{Spielman}}]{TopoColdAtoms_RMP}%
\BibitemOpen
\bibfield  {author} {\bibinfo {author} {\bibfnamefont {N.~R.}\ \bibnamefont
		{Cooper}}, \bibinfo {author} {\bibfnamefont {J.}~\bibnamefont {Dalibard}}, \
	and\ \bibinfo {author} {\bibfnamefont {I.~B.}\ \bibnamefont {Spielman}},\
}\href {\doibase 10.1103/RevModPhys.91.015005} {\bibfield  {journal}
{\bibinfo  {journal} {Rev. Mod. Phys.}\ }\textbf {\bibinfo {volume} {91}},\
\bibinfo {pages} {015005} (\bibinfo {year} {2019})}\BibitemShut {NoStop}%
\bibitem [{\citenamefont {Asb{\'o}th}\ \emph {et~al.}(2016)\citenamefont
	{Asb{\'o}th}, \citenamefont {Oroszl{\'a}ny},\ and\ \citenamefont
	{P{\'a}lyi}}]{asboth2016shortTI}%
\BibitemOpen
\bibfield  {author} {\bibinfo {author} {\bibfnamefont {J.~K.}\ \bibnamefont
		{Asb{\'o}th}}, \bibinfo {author} {\bibfnamefont {L.}~\bibnamefont
		{Oroszl{\'a}ny}}, \ and\ \bibinfo {author} {\bibfnamefont {A.}~\bibnamefont
		{P{\'a}lyi}},\ }\href@noop {} {\bibfield  {journal} {\bibinfo  {journal}
		{Lecture notes in physics}\ }\textbf {\bibinfo {volume} {919}} (\bibinfo
	{year} {2016})}\BibitemShut {NoStop}%
\bibitem [{\citenamefont {Trotter}(1959)}]{Trotter}%
\BibitemOpen
\bibfield  {author} {\bibinfo {author} {\bibfnamefont {H.}~\bibnamefont
		{Trotter}},\ }\href {\doibase 10.1090/S0002-9939-1959-0108732-6} {\bibfield
	{journal} {\bibinfo  {journal} {Proc. Amer. Math. Soc.}\ }\textbf {\bibinfo
		{volume} {10}},\ \bibinfo {pages} {545} (\bibinfo {year} {1959})}\BibitemShut
{NoStop}%
\bibitem [{\citenamefont {Suzuki}(1976)}]{Suzuki}%
\BibitemOpen
\bibfield  {author} {\bibinfo {author} {\bibfnamefont {M.}~\bibnamefont
		{Suzuki}},\ }\href {\doibase 10.1007/BF01609348} {\bibfield  {journal}
	{\bibinfo  {journal} {Communications in Mathematical Physics}\ }\textbf
	{\bibinfo {volume} {51}},\ \bibinfo {pages} {183} (\bibinfo {year}
	{1976})}\BibitemShut {NoStop}%
\bibitem [{\citenamefont {Hubbard}(1959)}]{continuous-HS-transformation}%
\BibitemOpen
\bibfield  {author} {\bibinfo {author} {\bibfnamefont {J.}~\bibnamefont
		{Hubbard}},\ }\href {\doibase 10.1103/PhysRevLett.3.77} {\bibfield  {journal}
	{\bibinfo  {journal} {Phys. Rev. Lett.}\ }\textbf {\bibinfo {volume} {3}},\
	\bibinfo {pages} {77} (\bibinfo {year} {1959})}\BibitemShut {NoStop}%
\bibitem [{\citenamefont {Hirsch}(1983)}]{HS_transform_discrete}%
\BibitemOpen
\bibfield  {author} {\bibinfo {author} {\bibfnamefont {J.~E.}\ \bibnamefont
		{Hirsch}},\ }\href {\doibase 10.1103/PhysRevB.28.4059} {\bibfield  {journal}
	{\bibinfo  {journal} {Phys. Rev. B}\ }\textbf {\bibinfo {volume} {28}},\
	\bibinfo {pages} {4059} (\bibinfo {year} {1983})}\BibitemShut {NoStop}%
\bibitem [{\citenamefont {Shi}\ and\ \citenamefont
	{Zhang}(2013)}]{SymmetryHao}%
\BibitemOpen
\bibfield  {author} {\bibinfo {author} {\bibfnamefont {H.}~\bibnamefont
		{Shi}}\ and\ \bibinfo {author} {\bibfnamefont {S.}~\bibnamefont {Zhang}},\
}\href {\doibase 10.1103/PhysRevB.88.125132} {\bibfield  {journal} {\bibinfo
	{journal} {Phys. Rev. B}\ }\textbf {\bibinfo {volume} {88}},\ \bibinfo
{pages} {125132} (\bibinfo {year} {2013})}\BibitemShut {NoStop}%
\bibitem [{\citenamefont {Shi}\ \emph {et~al.}(2015{\natexlab{b}})\citenamefont
	{Shi}, \citenamefont {Chiesa},\ and\ \citenamefont {Zhang}}]{2DFG_AFQMC}%
\BibitemOpen
\bibfield  {author} {\bibinfo {author} {\bibfnamefont {H.}~\bibnamefont
		{Shi}}, \bibinfo {author} {\bibfnamefont {S.}~\bibnamefont {Chiesa}}, \ and\
	\bibinfo {author} {\bibfnamefont {S.}~\bibnamefont {Zhang}},\ }\href
{\doibase 10.1103/PhysRevA.92.033603} {\bibfield  {journal} {\bibinfo
		{journal} {Phys. Rev. A}\ }\textbf {\bibinfo {volume} {92}},\ \bibinfo
	{pages} {033603} (\bibinfo {year} {2015}{\natexlab{b}})}\BibitemShut
{NoStop}%
\bibitem [{\citenamefont {Shi}\ and\ \citenamefont {Zhang}(2016)}]{inf_var}%
\BibitemOpen
\bibfield  {author} {\bibinfo {author} {\bibfnamefont {H.}~\bibnamefont
		{Shi}}\ and\ \bibinfo {author} {\bibfnamefont {S.}~\bibnamefont {Zhang}},\
}\href {\doibase 10.1103/PhysRevE.93.033303} {\bibfield  {journal} {\bibinfo
	{journal} {Phys. Rev. E}\ }\textbf {\bibinfo {volume} {93}},\ \bibinfo
{pages} {033303} (\bibinfo {year} {2016})}\BibitemShut {NoStop}%
\end{thebibliography}
\end{document}